\documentclass[prd,nofootinbib,preprint,superscriptaddress]{revtex4}

\usepackage{amsmath, amssymb, amsthm, graphicx, epsfig, fancyhdr,epsfig, slashed}

\usepackage{tikzsymbols}
\usepackage{natbib}
\usepackage{float}
\usepackage{xcolor}

\usepackage{enumitem}
\usepackage{amsmath}

\usepackage{tikz,xcolor,hyperref}

\usepackage{slashed}

\usepackage{subcaption}
\usepackage{braket}

\usepackage{hhline}
\usepackage{multirow}

\definecolor{lime}{HTML}{A6CE39}
\DeclareRobustCommand{\orcidicon}{
	\begin{tikzpicture}
	\draw[lime, fill=lime] (0,0) 
	circle [radius=0.2] 
	node[white] {{\fontfamily{qag}\selectfont \tiny ID}};
	\draw[white, fill=white] (-0.0625,0.095) 
	circle [radius=0.007];
	\end{tikzpicture}
	\hspace{-2mm}
}

\foreach \x in {A, ..., Z}{\expandafter\xdef\csname orcid\x\endcsname{\noexpand\href{https://orcid.org/\csname orcidauthor\x\endcsname}
			{\noexpand\orcidicon}}
}

\newcommand{\be}{\begin{equation}}
\newcommand{\ee}{\end{equation}}
\newcommand{\bea}{\begin{eqnarray}}
\newcommand{\eea}{\end{eqnarray}}

\newcommand{\ba}{\begin{eqnarray}}
\newcommand{\ea}{\end{eqnarray}}
\newcommand{\bi}{\begin{itemize}}
\newcommand{\ei}{\end{itemize}}

\newcommand{\x}{\star}

\def\s{\sigma}

\def\checkmark{\tikz\fill[scale=0.3, color=blue](0,.35) -- (.25,0) -- (1,.7) -- (.25,.15) -- cycle;}

\begin{document}

\title{Inflationary Gravitational Wave Spectral Shapes \\ as test for Low-Scale Leptogenesis}

\author{Zafri A. Borboruah}
\email{zafri123@iitb.ac.in}
\affiliation{Indian Institute of Technology Bombay, Mumbai 400076, Maharashtra, India}

\author{Anish Ghoshal}
\email{anish.ghoshal@fuw.edu.pl}
\affiliation{Institute of Theoretical Physics, Faculty of Physics,\\ University of Warsaw,
ul. Pasteura 5, 02-093 Warsaw, Poland}

\author{Lekhika Malhotra}
\email{lekhika.malhotra@iitb.ac.in}
\affiliation{Indian Institute of Technology Bombay, Mumbai 400076, Maharashtra, India}

\author{Urjit Yajnik}
\email{yajnik@iitb.ac.in}
\affiliation{Indian Institute of Technology Gandhinagar, Gandhinagar 382055, India}
\affiliation{Indian Institute of Technology Bombay, Mumbai 400076, Maharashtra, India}

\begin{abstract}
We study non-thermal resonant leptogenesis in a general setting where a heavy majoron $\phi$ decays to right-handed neutrinos (RHNs) whose further out-of-equilibrium decay generates the required lepton asymmetry. Domination of the energy budget of the Universe by the $\phi$ or the RHNs alters the evolution history of the primordial gravitational waves (PGW) of inflationary origin, which re-enter the horizon after inflation, modifying the spectral shape. The decays of $\phi$ and RHNs release entropy into the early Universe while nearly degenerate RHNs facilitate low and intermediate-scale leptogenesis. A characteristic damping of the GW spectrum resulting in knee-like features would provide evidence for low-scale non-thermal leptogenesis. We explore the parameter space for the lightest right-handed neutrino mass $M_1\in[10^2,10^{14}]$ GeV and washout parameter $K$ that depends on the light-heavy neutrino Yukawa couplings $\lambda$, 
in the weak ($K < 1$) and strong ($K > 1$) washout regimes. The resulting novel features compatible with observed baryon asymmetry are detectable by future experiments like LISA and ET. By estimating signal-to-noise ratio (SNR) for upcoming GW experiments, we investigate the effect of the majoron mass $M_\phi$ and reheating temperature $T_\phi$, which depends on the $\phi-N$ Yukawa couplings $y_N$. 
\end{abstract}

\maketitle
\tableofcontents
\flushbottom

\section{Introduction}\label{sec:intro}
The Standard Model (SM) of particle physics cannot explain tiny neutrino masses and the observed matter-antimatter asymmetry of the Universe. Neutrino oscillation experiments involving solar \cite{Super-Kamiokande:2001bfk,Super-Kamiokande:2002ujc,SNO:2002tuh,Super-Kamiokande:2005mbp,Super-Kamiokande:2016yck,Borexino:2015axw}, atmospheric \cite{IceCube:2017lak, ANTARES:2018rtf} and reactor \cite{KamLAND:2008dgz,T2K:2011ypd,DoubleChooz:2011ymz,T2K:2013ppw} neutrinos provide evidence that neutrinos 
have mass and the flavor states mix due to the propagation of multiple mass eigenstates. This however only quantifies the squared mass differences of the neutrinos and not the absolute mass scales involved. On the other hand, the $\beta$-decay experiment KATRIN \cite{KATRIN:2021uub} gives a stringent direct limit on the absolute value of neutrino mass scale, $m_\nu < 0.8 $ eV.

On the cosmological frontiers, measurements of the Cosmic Microwave Background radiation (CMBR) by Planck 2018 \cite{Planck:2018vyg} and large-scale structure (LSS) constrain the sum of all neutrino masses to $\sum_i m_{\nu_i} < 0.12$ eV \cite{Aghanim:2018eyx,eBOSS:2020yzd}. On the other hand, the observed baryon asymmetry of the Universe (BAU) \cite{Zyla:2020zbs,Aghanim:2018eyx}, often expressed in terms of the  baryon to photon ratio is, according to the Planck data~\cite{Aghanim:2018eyx}, 
\begin{equation}
\eta_B^{\rm CMB} = \frac{n_{B}-n_{\bar{B}}}{n_{\gamma}} = (6.21 \pm 0.16) \times 10^{-10}
\label{eq:eta}
\end{equation}
This agrees with the values of abundances of light elements extracted from BBN  (Big Bang Nucleosynthesis) data~\cite{Fields:2019pfx}.

Just augmenting the Standard Model (SM) with two or more right-handed Majorana neutrinos (RHNs), one can explain the tiny neutrino masses, generated via the well-known Type-I seesaw mechanism \cite{Minkowski:1977sc,Yanagida:1979as,Glashow:1979nm,Mohapatra:1979ia}, 
while baryon asymmetry can be explained using the baryogenesis via leptogenesis mechanism~\cite{Fukugita:1986hr}, everything tied together under one umbrella. However for this mechanism of baryogenesis to be successful, the lightest RHN mass is required to exceed the Davidson-Ibarra bound, $M_1 \gtrsim 10^9$ GeV, assuming the RHN mass spectrum is hierarchical, and also assuming no lepton flavor effects~\cite{Nardi:2006fx,Abada:2006fw,Abada:2006ea,Davidson:2002qv,Giudice:2003jh}. This destroys any hope of direct detection of the high-scale seesaw and leptogenesis as they lie well beyond the energy reach of the current or foreseeable future laboratory experiments or astrophysical observations making it impossible to test such new physics. 

Nonetheless, certain indirect signatures of new physics like lepton number violation processes through neutrinoless double beta decay~\cite{Cirigliano:2022oqy} or CP violation in neutrino oscillation~\cite{Endoh:2002wm} are searched for in laboratories. 
On the other hand, theoretical constraints on the low energy values of couplings, consistent with UV-completion such as $SO(10)$ Grand Unified Theories (GUT) ~\cite{DiBari:2008mp,Bertuzzo:2010et,Buccella:2012kc,Altarelli:2013aqa,Fong:2014gea,Mummidi:2021anm,Patel:2022xxu} or with the requirement of electroweak (EW) vacuum (meta)stability in the early Universe~\cite{Ipek:2018sai,Croon:2019dfw} allow us to pose some bounds on the large parameter space involved in seesaw and leptogenesis. 

The discovery of Gravitational Waves (GWs) from black hole mergers by LIGO and Virgo collaboration \cite{LIGOScientific:2016aoc,LIGOScientific:2016sjg} and evidence for measurements of stochastic GW background by pulsar timing array (PTA) observations~\cite{Carilli:2004nx,Janssen:2014dka,Weltman:2018zrl,EPTA:2015qep,EPTA:2015gke,NANOGrav:2023gor,NANOGrav:2023hvm} have lead to several new physics cases to detect GWs of primordial origins as well. Particularly in the context of baryogenesis via leptogenesis, cosmological pathways to probe such high-scale physics involve predictions of the CMB spectral indices~\cite{Ghoshal:2022fud} or gravitational waves from local cosmic strings~\cite{Dror:2019syi, Saad:2022mzu, DiBari:2023mwu, Blasi:2020wpy}, global cosmic strings \cite{Fu:2023nrn}, domain walls~\cite{Barman:2022yos, King:2023cgv}, nucleating and colliding vacuum bubbles~\cite{Dasgupta:2022isg,Borah:2022cdx}, other topological defects~\cite{Dunsky:2021tih}, graviton bremmstrahlung~\cite{Ghoshal:2022kqp}, gravitational production of neutrinos~\cite{Haque:2023zhb}, inflationary tensor perturbations propagating as GW~\cite{Berbig:2023yyy,Borboruah:2024eal}, primordial blackholes~\cite{Perez-Gonzalez:2020vnz,Datta:2020bht,Barman:2024slw,JyotiDas:2021shi,Barman:2021ost,Bernal:2022pue,Bhaumik:2022pil,Bhaumik:2022zdd,Ghoshal:2023fno}, and exploring the oscillatory features of the primordial non-gaussianity and curvature bi-spectrum and tri-spectrum \cite{Cui:2021iie,Fong:2023egk}.

Under this circumstance, we need more such probes of BSM physics and form a synergy between cosmological observations and laboratory experiments involving seesaw and low-scale leptogenesis. In this paper, we take a first step in this direction and investigate a general scenario where a heavy scalar called majoron, denoted by $\phi$, may decay to create an initial non-thermal abundance of right-handed neutrinos (RHNs) whose subsequent decay leads to leptogenesis satisfying Eq.~\eqref{eq:eta}. This majoron $\phi$ is different from inflaton. Inflaton itself is not the direct concern of this work although we use the inflationary reheating temperature $T_{\rm RH}$ in our calculations. We will be concerned with gravitational waves that are generated due to quantum tensor fluctuations of the metric during inflation~\cite{Grishchuk:1974ny,Starobinsky:1979ty,Rubakov:1982df,Haque:2021dha,Guzzetti:2016mkm}. These GWs can act as a book-keeping of the expansion history of our Universe~\cite{Seto:2003kc,Boyle:2005se,Boyle:2007zx,Kuroyanagi:2008ye,Nakayama:2009ce,Kuroyanagi:2013ns,Jinno:2013xqa,Saikawa:2018rcs,Haque:2023zhb,Chen:2024roo}.
Since the detailed time evolution of the Hubble rate during
the expansion determines various PGW frequencies that are red-shifted to the present day, thus allowing us to investigate standard and non-standard thermal history of the universe in the pre-BBN era~\cite{Bernal:2020ywq,Nakayama:2008ip,Nakayama:2008wy,Kuroyanagi:2011fy,Buchmuller:2013lra,Buchmuller:2013dja,Jinno:2014qka,Kuroyanagi:2014qza,Haque:2021dha,Ghoshal:2023sfa,Ghoshal:2022ruy,Chen:2024roo}.
In this context, since we are interested in low-scale seesaw and leptogenesis, we will study a scenario where $\phi$ dominates the energy budget of the Universe which leads to interesting GW spectral shapes. On one hand, $\phi$ decay will facilitate low-scale leptogenesis and on the other hand, it is also responsible for the generation of GW spectral shapes that can be measured by various GW detectors. In particular, we consider an epoch of intermediate matter domination by $\phi$ particles affecting the post-inflationary evolution of the Universe~\cite{Scherrer:1984fd,Barman:2024slw,Kolb:1990vq,Bezrukov:2009th}.

In this model, non-thermal leptogenesis is achieved via the majoron decay to RHNs. The leptogenesis scale is brought down mainly by the resonance effect that enhances the CP parameter $\epsilon$. RHNs are produced in an excess initial abundance after the majoron has decayed, compared to the thermal case. Depending on reheating temperature $T_\phi$ and washout parameter $K$, the final efficiency factor $\kappa_f$ may get enhanced compared to the thermal case due to this excess initial abundance~\cite{Hahn-Woernle:2008tsk,Zhang:2023oyo}. Unlike~\cite{Berbig:2023yyy}, here the RHNs need not dominate the Universe since the majoron is assumed to dominate the energy density of the Universe which distorts inflationary GWs. Therefore we do not need to stick to only the $K\ll1$ condition of~\cite{Berbig:2023yyy} for RHN domination. Note that RHNs can still dominate after $\phi$ decays, depending on parameters.

The decays occur far away from thermal equilibrium, therefore it is bound to release a large amount of entropy dump, which in turn dilutes the energy density of primordial GWs that enter the horizon before the decay. We show that a 2-step entropy injection scenario is possible if the RHNs produced are highly relativistic at production, which eventually become non-relativistic and then decay at a lower temperature. In this case, two knee-like features occur in the GW spectrum due to the majoron and RHN decays which we show for the first time. We will show the capability of several upcoming GW detectors like LISA, DECIGO, BBO, etc. in probing the parameter space for such leptogenesis, particularly in regions where it is impossible to test via laboratory or astrophysical experiments.

 In this work a framework of non-thermal \emph{resonant} leptogenesis from the decay chain $\phi \to N_{1,2} \to H+L$, includes entropy release from both $\phi$ and RHNs, spans RHN masses $10^2$–$10^{14}$ GeV and washout strengths $K\sim10^{-20}$–$10^4$, systematically classifies scenarios, and highlights knee-like damping and kinked GW spectra with SNR estimates across detectors. Unlike the above studies, we focus on a nearly-degenerate RHN sector decoupled from the thermal bath, providing a general yet distinctive link between majoron-induced resonant leptogenesis and observable GW features\footnote{See Refs.~\cite{DEramo:2019tit,Datta:2022tab,Chianese:2024nyw,Samanta:2025jec} for ideas related to probing non-standard expansion histories and leptogenesis through inflationary gravitational waves.}.

\textit{The paper is organized as follows:} In Sec.~\ref{sec:thermal_lg}, we give a brief review of the thermal leptogenesis mechanism and the Davidson-Ibarra bound on heavy neutrino masses. We consider resonant leptogenesis to bring the leptogenesis scale down. In Sec.~\ref{sec:lepscalar}, we discuss the majoron model that facilitates non-thermal leptogenesis in details and the dynamics of majoron and RHN decays. We construct the relevant Boltzmann equations. In Sec.~\ref{sec:classification}, we identify three classes of scenarios relevant to GW observations. We provide analytical approximations of matter-radiation equilibrium temperature and dilution factor from entropy injection for these classifications. Then in Sec.~\ref{sec:distortion} we discuss primordial gravitational waves from inflation with and without an intermediate matter domination epoch. We also introduce a transfer function to encapsulate a 2-step entropy injection. This section is followed by our results in Sec.~\ref{sec:results}. Finally, we conclude this paper in Sec.~\ref{sec:conclusion}. In Appendix~\ref{app:C}, we explicitly demonstrate a scenario where washout factor can be tiny while keeping the CP asymmetry large. In Appendix~\ref{app:A} we describe the Boltzmann equations in terms of comoving quantities.

\medskip

\section{Review of Thermal Leptogenesis}
\label{sec:thermal_lg}
Before discussing the effect of majoron field reheating on leptogenesis, let us first give a brief review of standard thermal leptogenesis. Initially proposed by Fukugita and Yanagida in 1986~\cite{Fukugita:1986hr, Yanagida:2005hz}, thermal leptogenesis hinges on the out-of-equilibrium decay of RHNs, which are thermally produced through scattering processes in the early Universe from the thermal bath of the primordial SM plasma. The Lagrangian is given by,
\begin{equation}\label{eq:Lthermal}
    \mathcal{L}_{\rm thermal}={\cal L}_{\rm SM} 
	+ i\overline{N} \slashed{\partial}N
	- \left(\lambda \overline{L} \tilde{H} N
	+ \frac{M_N}{2} \overline{N^C} N + \text{h.c} \right)
\end{equation}
where $\mathcal{L}_{\rm SM}$ is the SM Lagrangian, $N$ represents RHNs, $\tilde{H}=i\sigma_2H^*$ where $H$ is the SM Higgs doublet and $L$ is the SM leptonic doublet. Here $M_N=\text{diag}[M_1,M_2,M_3]$ is the RHN Majorana mass matrix and $\lambda$ encompasses the light-heavy Yukawa couplings. Due to the Majorana mass term of RHNs, after electroweak symmetry breaking (EWSB), the familiar Type-I seesaw mechanism \cite{Yanagida:1979as} provides mass to the light neutrinos, $m_\nu=v^2 \lambda^T M_N^{-1} \lambda$, where $v\sim174$ GeV is the vacuum expectation value (VEV) of the SM Higgs. On the other hand, RHNs of Majorana nature can decay into both leptons and anti-leptons through the Yukawa coupling matrix $\lambda$, thereby violating lepton number (LNV),
 \begin{equation}
 N_1\rightarrow H + L \quad \text{and}\quad N_1\rightarrow H^\dag + L^\dag
 \end{equation}
 
Typically, this scenario assumes a hierarchical heavy neutrino mass spectrum, where the masses of the heavier RH neutrinos, $M_2$ and $M_3$, are much greater than the lightest RHN mass, $M_1$. This condition allows the lightest neutrino, $N_1$ to generate the necessary lepton asymmetry when CP is violated in their decay, after washing out any asymmetry created by $N_2$ and $N_3$. The CP asymmetry, denoted as $\epsilon_1$, is given by:
\begin{equation}
    \label{eq:ep1}
    \epsilon_1=\frac{\Gamma_{N_1\rightarrow HL}-\Gamma_{N_1\rightarrow H^\dag L^\dag}}{\Gamma_{N_1}^{rf}}
\end{equation}
where $\Gamma_{N_1\rightarrow Hl}$ is the decay width of the $N_1\rightarrow H+l$ decay and 
\begin{equation}
    \Gamma_{N_1}^{rf}=\Gamma_{N_1\rightarrow HL}+\Gamma_{N_1\rightarrow H^\dag L^\dag}=\frac{|\lambda\lambda^\dag|_{11}}{8\pi}M_1
\end{equation}
is the total decay width of $N_1$ in the rest frame of the neutrino.
The lepton asymmetry can be re-written as $B-L$ asymmetry denoted by the number density $n_{B-L}$, which is subsequently converted into the baryon asymmetry through $B+L$ conserving electroweak sphaleron processes~\cite{Harvey:1990qw,Khlebnikov:1988sr,Luty:1992un}. When there is no initial $B-L$ asymmetry, the final baryon asymmetry, denoted as baryon-to-photon number density ratio $\eta_B$, is expressed as~\cite{Buchmuller:2004nz},
\begin{equation}
    \label{eq:etaB}
    \eta_B\equiv\frac{n_B-n_{\overline{B}}}{n_\gamma}=\frac{3}{4}\frac{a_\text{sph}}{f}\,\epsilon_1\kappa_f \simeq 0.96\times10^{-2}\,\epsilon_1\kappa_f
\end{equation}
where $a_\text{sph}=28/79$~\cite{Arnold:1987mh,Khlebnikov:1988sr,Harvey:1990qw} is the efficiency of the sphaleron processes of converting $B-L$ asymmetry into baryon asymmetry and $f=2387/86$ is the dilution factor due to production of photons since the onset of leptogenesis till recombination assuming an isentropic expansion of the Universe. And $\kappa_f$ is the \textit{efficiency parameter} \cite{Buchmuller:2004nz}, which is independent of $\epsilon_1$ and which parameterizes the efficiency of lepton asymmetry production depending on the initial $N_1$ abundance and washout. It can be calculated by solving the relevant Boltzmann equations governing the evolution of the $N_1$ abundance and $B-L$ asymmetry~\cite{Nardi:2007jp,Basboll:2006yx}. Its final value is normalized to unity if a thermal initial abundance of $N_1$ is assumed and there is no washout. 

\subsection{Conditions for Departure from Thermal Equilibrium}
One of the key conditions for successful baryogenesis, as outlined by Sakharov, is the departure from thermal equilibrium. In the context of leptogenesis, this condition is satisfied if the total decay rate of $N_1$ is smaller than the Hubble expansion rate of the Universe when the temperature is approximately equal to the mass of $N_1$:
\begin{equation}
    \label{eq:out-of-eq-1}
    \Gamma_{N_1}^{rf}<\mathcal{H}(T\sim M_1)
\end{equation}
where $\mathcal{H}(T\sim M_1)\equiv\mathcal{H}(M_1)=\sqrt{\frac{8\pi^3g_\ast}{90}}\frac{M_1^2}{M_\text{pl}}$ is the Hubble parameter with $M_\text{pl}=1.22\times10^{19}$ GeV and $g_*$ is the relativistic degrees of freedom which is approximately equal to $106.75$ at the high temperatures under discussion. It can be shown that the above condition is satisfied when the \textit{effective neutrino mass}, defined by,

\begin{equation}
    \label{eq:effneutrinomass}
    \tilde{m}_1=\frac{(m_D m_D^\dag)_{11}}{M_1}
\end{equation}

where $m_D=\lambda\, v$ is the Dirac mass matrix with $v=\braket{H}\simeq174$ GeV being the vacuum expectation value (VEV) of the Standard Model Higgs, is smaller than the \textit{equilibrium neutrino mass}, defined as,
\begin{equation}
    \label{eq:eqneutrinomass}
    m_\ast=\frac{16\pi^{5/2}}{3\sqrt{5}}g_*^{1/2}\frac{v^2}{M_{\rm pl}}\simeq 1.1\times 10^{-3}\text{eV}
\end{equation}
The ratio is called the decay parameter or the washout parameter,
\begin{equation}
    \label{eq:Kdefinition}
    K\equiv\frac{\Gamma_{N_1}^{r f}}{\mathcal{H}(M_1)}=\frac{\tilde{m}_1}{m_\ast}
\end{equation}
which determines whether $N_1$ decays are in thermal equilibrium or not. The case $K<1$ is called the \textit{weak washout regime}, where Yukawa interactions of $N_1$ with light neutrinos via the SM Higgs are weak, hence washout of any initial asymmetry via reverse processes is small. Therefore leptogenesis is unavoidable in this regime. However in the so-called \textit{strong washout regime} ($K>1$), although Yukawa interactions are strong, leptogenesis can still occur and the final asymmetry depends on the interplay between decay, inverse decay, and scattering of RHNs~\cite{Buchmuller:2004nz}.

Within type-I seesaw framework, the Yukawa couplings $\lambda$ can be expressed by the Casas-Ibarra parametrization~\cite{Casas:2001sr},
\begin{equation}\label{eq:casas}
    \lambda=\frac{1}{v}\, M_N^{1/2}\,R\,m_\nu^{1/2}U_{\rm PMNS}^\dag,
\end{equation}
where $m_\nu=\text{diag}(m_1,m_2,m_3)$, $M_N=\text{diag}(M_1,M_2,M_3)$ and $R$ is a complex orthogonal matrix. The effective neutrino masses, 
\begin{equation}\label{eq:mtilde2}
\tilde{m}_i=\frac{\left|\lambda\lambda^\dag\right|_{ii}v^2}{M_i}=\sum_jm_j|R_{ij}|^2
\end{equation}
correspond to the physical neutrino masses $m_i$ if $R$ is diagonal. 
This means, assuming normal ordering $m_1<m_2<m_3$ and $m_1=0$, for small off-diagonal elements of $R$ matrix, the effective neutrino mass $\tilde{m}_1\rightarrow0$, hence the washout parameter $K\rightarrow 0$. However, in this case, $\tilde{m}_2\rightarrow10^{-3}$ eV since $m_2=\sqrt{m_1^2+\Delta m_{\rm atm}^2}\sim8.6\times10^{-3}$ eV, implying the washout factor $K_2=\tilde{m}_2/m_*$ for $N_2$ remains $\mathcal{O}(1)$. Similarly $K_3$ also remains large. It can be shown that, in general, $\tilde{m}_1$ is larger than the lightest active neutrino mass~\cite{Fujii:2002jw},
\begin{equation}
    \tilde{m}_1>\text{Min}[m_1,m_2,m_3]
\end{equation}

It turns out the final efficiency parameter $\kappa_f$ can be estimated analytically for both weak and strong washout regimes~\cite {Buchmuller:2004nz}. In the strong washout regime, $\kappa_f$ is estimated by,
\begin{equation}
    \label{eq:kappaf_strong}
    \kappa_f(K)\simeq(2\pm1)\times10^{-2}\left(\frac{0.01\text{ eV}}{m_*\,K}\right)^{1.1\pm0.1}
\end{equation}
while in the weak washout regime, $\kappa_f$ depends on the initial abundance of $N_1$. For thermal initial abundance, it is approximately 1 in the weak regime while for vanishing initial abundance, it is estimated as,
\begin{equation}
    \label{eq:kappaf_weak_vanishing}
    \kappa_f(K)\simeq\frac{9\pi^2K^2}{64},\quad(K<1)
\end{equation}
For thermal initial $N_1$ abundance, $\kappa_f$ can be generically written for all $K$ as,
\begin{align}
    \label{eq:kappaf_thermal}
    \kappa_f(K)&\simeq\frac{2}{z_B(K)K}\left(1-e^{\frac{1}{2}z_B(K)K}\right)\\
    \label{eq:zB}
    \text{where }z_B(K)&\simeq1+\frac{1}{2}\text{ ln}\left(1+\frac{\pi K^2}{1024}\left[\text{ ln}\left(\frac{3125\pi K^2}{1024}\right)\right]^5\right)
\end{align}

In Fig.~\ref{fig:kappaThermal} we show $\kappa_f$ as a function of washout parameter $K$ for thermal (red solid line) and vanishing (blue dashed line) initial abundance of $N_1$. Depending on the CP parameter $\epsilon_1$ and $K$, the final baryon asymmetry can be calculated from Eq.~\eqref{eq:etaB}.

\subsection{CP Asymmetry}
\begin{figure}[!ht]
\centering
\includegraphics[width=0.8\linewidth]{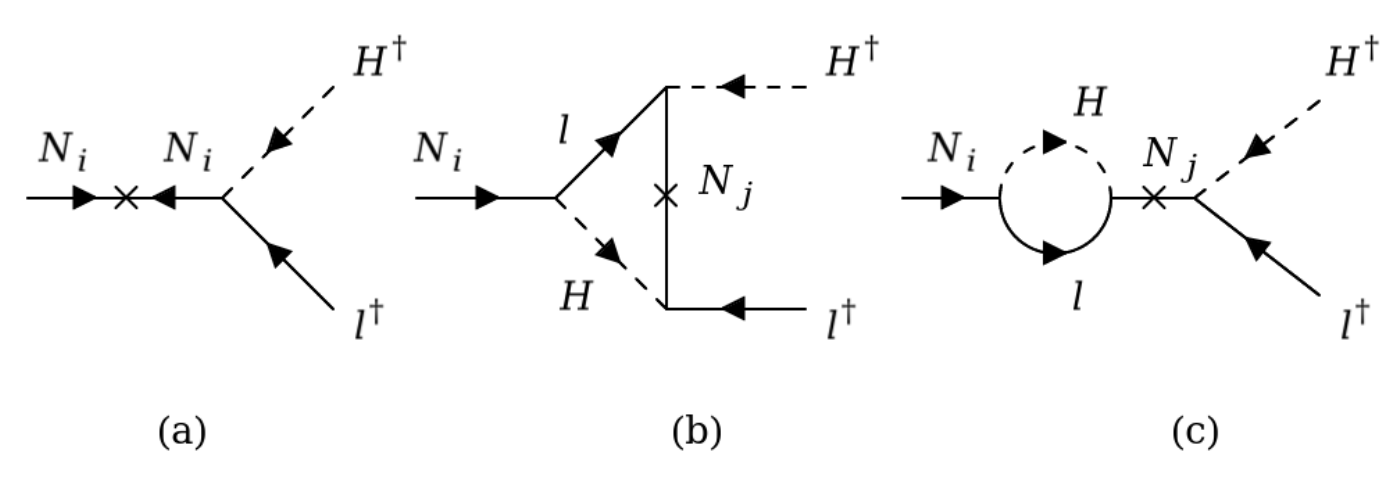}  
\caption{\small \it Feynmann diagrams of $N\rightarrow lH$ decay process at the (a) tree-level and (b)-(c) 1-loop level. (b) represents the vertex contribution and (c) represents the self-energy contribution.}
    \label{fig:feynmann}
\end{figure}

The CP asymmetry $\epsilon_1$ is determined by the interference between the tree level and 1-loop diagrams ~\cite{Flanz:1994yx,Covi:1996wh} and can be calculated from $lH\rightarrow lH$ scattering processes~\cite{Buchmuller:1997yu}. The relevant Feynmann diagrams giving tree level, vertex, and self-energy contributions are shown in Fig.~\ref{fig:feynmann}. 

For RHN mass hierarchy $M_1<M_2<M_3$, the CP parameter is obtained as,
\begin{equation}\label{eq:epsilon2}
    |\epsilon_1|=\sum_{i\neq1}\frac{3}{16\pi}\frac{M_1}{M_i}\frac{\text{Im}\left[\left(\lambda\lambda^\dag\right)_{1i}^2\right]}{|\lambda\lambda^\dag|_{11}}=\frac{3}{16\pi}\frac{M_1}{v^2}\frac{\sum_im_i^2\text{Im}\left[R_{1i}^2\right]}{\sum_jm_j|R_{1j}|^2}<\epsilon_1^{\rm max}
\end{equation}

where the maximal asymmetry is given by~\cite{Buchmuller:2003gz,Hahn-Woernle:2008tsk,Buchmuller:2004tu,Davidson:2002qv,Hambye:2003rt},
\begin{equation}
    \label{eq:ep1max}
    \epsilon_1^\text{max}=\frac{3}{16\pi}\frac{M_1}{v^2}(m_3-m_1)\simeq 10^{-6}\left(\frac{M_1}{10^{10}\text{ GeV}} \right)\left(\frac{m_3-m_1}{0.05 \text{ eV}}\right)
\end{equation}
Note that for small values of $\tilde{m}_1$ we require small values of $|R_{1i}|^2$ as seen from Eq.~\eqref{eq:mtilde2}. However, that does not necessarily make $|\epsilon_1|$ tiny as it depends on the ratio of off-diagonal elements of $R$ as seen in Eq.~\eqref{eq:epsilon2}. It can be verified that it is possible to achieve maximal CP asymmetry even if $\tilde{m}_1\rightarrow0$, implying $K\rightarrow0$ (see Appendix~\ref{app:C}). For our analysis, we will consider the range $K\in[10^{-20},10^4]$ while keeping $K_2,K_3>1$. 

\subsection{Lower Bounds on Heavy Neutrino Masses}
Putting the maximal CP asymmetry parameter $\epsilon_1^\text{max}$ in Eq. \eqref{eq:etaB} gives the maximal baryon asymmetry $\eta_B^\text{max}$, which needs to be larger than the observed asymmetry $\eta_B^\text{CMB}$, i.e. $\eta_B^\text{max}\geq \eta_B^\text{CMB}$. Then one finds the Davidson-Ibarra bound on $M_1$ using Eq.~\eqref{eq:etaB}~\cite{Davidson:2002qv,Buchmuller:2002rq,Buchmuller:2004nz},
\begin{equation}
    \label{eq:M1lower}
    M_1\geq 6.4\times10^8\text{ GeV}\left(\frac{\eta_B^\text{CMB}}{6\times 10^{-10}}\right)\left(\frac{0.05 \text{ eV}}{m_3-m_1}\right)\kappa_f^{-1}
\end{equation}
Assuming the final efficiency factor $\kappa_f=1$ gives the $3\sigma$ bound on the minimum reheating temperature $T_{\rm RH}$ at the end of  inflation~\cite{Buchmuller:2004tu},
\begin{equation}
    \label{eq:M1lower3sig}
    T_\text{RH}\gtrsim M_1\gtrsim 4\times10^{8}\text{ GeV}
\end{equation}
This stems from the condition that $T_{\rm RH}$ needs to be larger than the initial temperature of leptogenesis $T_i\sim M_1$ when the RHN decay process becomes out-of-equilibrium (see Eq.~\eqref{eq:out-of-eq-1})~\cite{Buchmuller:2005eh}.


\subsection{Resonant Leptogenesis}\label{sec:resonant}

Thermal leptogenesis operates at high temperatures, which in turn pushes the inflationary reheating temperature $T_{\rm RH}$ to a high scale, as shown in Eq.~\eqref{eq:M1lower} and~\eqref{eq:M1lower3sig}. Such high scales involving RHN masses $M_i\gtrsim10^9$ GeV are difficult to test in the lab. Also, a high neutrino mass conflicts with the Electroweak naturalness condition which requires RHN masses $M_i\lesssim10^7$ GeV~\cite{Vissani:1997ys,Clarke:2015gwa,Datta:2022tab}. A lower reheating temperature from inflation $T_{\rm RH}$ is motivated in different inflationary scenarios like~\cite{Drees:2021wgd,Ghoshal:2022zwu}. Even in some SUGRA models~\cite{Kawasaki:2004qu}, reheating temperature is found to be as low as $10^6$ GeV which contradicts a high $T_\text{RH}$, favoring low-scale-leptogenesis. Motivated by these reasons, and for lab testability, the scale of leptogenesis can be brought down by several orders in the so-called {\it resonant} regime which focuses on nearly degenerate heavy neutrinos leading to pronounced CP violation in their decays. It can be shown that if the mass difference between the two lightest right-handed neutrinos is of the order of the decay widths of them, i.e. $M_2 - M_1 \sim \Gamma_{N_2}^{r f}$ the CP parameter $\epsilon_1$ is enhanced via the self-energy contribution~\cite{Pilaftsis:1997dr,Hambye:2003rt} in the loop level diagrams shown in Fig~\ref{fig:feynmann}. 

The maximum CP asymmetry with resonant effect due to quasi-degenerate right-handed neutrinos $N_1$ and $N_2$ can be written as \cite{Berbig:2023yyy, Hambye:2003rt},
\begin{equation}
\label{eq:epsilon}
    \epsilon = \text{Min}\left[\epsilon_1^\text{max}\frac{S_2\cdot m_3-m_1}{m_3-m_1},1\right],\quad \text{where } S_2=\frac{M_2}{2\Gamma_{N_2}^{r f}} \quad\text{and } M_2-M_1=\frac{\Gamma_{N_2}^{r f}}{2}
\end{equation}
where, $\epsilon_1^{\rm max}$ is the maximal asymmetry in non-resonant case given by Eq.~\eqref{eq:ep1max}. It was shown that within the resonant regime, the lowest RHN mass $M_1$ can be as low as 1 TeV~\cite{Pilaftsis:1997jf}. In~\cite{Pilaftsis:2005rv} this scale is shown to reach upto the electroweak symmetry breaking scale $\sim100$ GeV, below which sphaleron processes are suppressed and the lepton asymmetry can not be effectively converted to baryon asymmetry. From here onwards we will consider resonant leptogenesis with $M_1\gtrsim 100$ GeV for our analysis. Note that in some alternate leptogenesis scenarios, e.g. via neutrino oscillations, BAU can be generated for even for RHN mass values $\sim100$ MeV~\cite{Klaric:2020phc}.

\medskip

\section{$B-L$ Symmetry Breaking and Non-thermal Leptogenesis}
\label{sec:lepscalar}

\par In this section, we construct a global $U(1)_{B-L}$ neutrino mass model inspired by~\cite{Fong:2023egk}, that includes all the essential components for non-thermal leptogenesis. In non-thermal leptogenesis, RHNs are produced from inflaton decay~\cite{Asaka:1999yd,Giudice:2003jh, Hahn-Woernle:2008tsk, Zhang:2023oyo} or from a generic scalar decay~\cite{Datta:2022tab,Nemevsek:2023yjl}. Such a scenario can lead to an excess initial abundance of RHNs compared to their thermal abundance. As a result, the final efficiency factor, $\kappa_f$ can be enhanced by a factor 100 which relaxes the Davidson-Ibarra bound on RHN mass given in Eq.~\ref{eq:M1lower}. In this paper, the model extends the SM to include right handed neutrinos $N_i$ where $i\in\{1,2,3\}$, and an additional scalar boson denoted as $\sigma$, with $B-L$ charge of $+2$. This complex scalar is responsible for generating the heavy Majorana masses of RHNs after spontaneously breaking the $U(1)_{B-L}$ symmetry. The Lagrangian is as follow,
\begin{align} \label{eq:Lagrangian}
   \nonumber {\cal L} 
	 = 
	{\cal L}_{SM} &
	+ i\overline{N} \slashed{\cal D}N
	+ |\mathcal{D}_\mu\sigma|^2
	- \left(\lambda \overline{L} \tilde{H} N
	+ \frac{y_N}{2} \sigma \overline{N^C} N + \text{h.c} \right)
	- V(H,\sigma)\,,\\
	V(H,\sigma)
	& = 
	\lambda_H (|H|^2 - v^2)^2 + \lambda_\sigma (|\sigma|^2 - v_{B-L}^2)^2 + \lambda_{\sigma H} |\sigma|^2 |H|^2\,,
\end{align}
where  $L$ is the leptonic doublet, $H$ denotes the Higgs doublet, and $\tilde{H}=i\sigma_2H^*$ as in Eq.~\eqref{eq:Lthermal}. The symbols $\lambda$, $y_N$, $\lambda_H$,  $\lambda_{\s H}$, and $\lambda_{\sigma}$ encompass the dimensionless coupling constants in the theory. The Yukawa coupling matrices $\lambda$ and $y_N=\text{diag}[\![y_{N_i}]\!]$  parameterize the strengths of the SM Higgs and the scalar couplings to the RHNs. Here $v_{B-L}$ is the $B-L$ symmetry-breaking scale. These parameters, as we will see, play key roles in determining the GW spectral shapes which we explore in details. Around the ${B-L}$ breaking vacuum, $\sigma$ can be parameterized as,
\begin{eqnarray}
	\sigma &=& (v_{B-L} + \rho)e^{i\phi/v_{B-L}}\,,
\end{eqnarray}
where $\phi$ is the Nambu-Goldstone (NG) boson called majoron. After $B-L$ symmetry is broken, the $N_i$ acquires the Majorana mass $M_i=y_{N_i} v_{B-L}$ where the extra phase is rotated away by redefining all fermions as $f\rightarrow f'=fe^{iq_f^{B-L}\phi/2v_{B-L}}$, where $q_f^{B-L}$ is the $B-L$ charge of the fermion $f$. The radial component $\rho$ acquires a mass $M_{\rho} = \sqrt{8\lambda_{\sigma}} v_{B-L}$ in the vanishing portal coupling limit, $ \lambda_{\sigma H} =0$. The majoron $\phi$ couples to the RHNs through its kinetic term which gives,
\begin{equation}
    \mathcal{L}\supset-\frac{\partial_\mu\phi}{2v_{B-L}}J^\mu_{B-L}=\frac{\phi}{2v_{B-L}}\partial_\mu J^\mu_{B-L},
\end{equation}
where the $B-L$ current is $J^\mu_{B-L}=\sum_f q^{B-L}_f\overline{f}\gamma^\mu f$. The second part of the equation is obtained by carrying out the integration by parts in the action and ignoring the surface term. This equation means that the majoron couples to $B-L$ violating currents, which in the RHN sector is the Majorana mass term of the RHNs. Hence, we obtain the majoron-RHN coupling,
\begin{equation}\label{eq:majoron RHN coupling}
    \mathcal{L}\supset\frac{i\phi}{v_{B-L}}M_i\overline{N^c_i}N_i.
\end{equation}
At this point, the majoron is massless. To give it mass, we consider a asymptotically free hidden gauge sector with confinement scale $\Lambda< v_{B-L}$, and make it anomalous under the $U(1)_{B-L}$, by adding fermions which are vector-like under the SM but chiral under the $U(1)_{B-L}$ and charged under the hidden group. This group becomes non-perturbative at the scale $\Lambda$, and the majoron couples to the new gauge bosons as follows,
\begin{equation}\label{eq:majoron hidden coupling}
    \mathcal{L}\supset\frac{\phi}{v_{B-L}}\frac{g_X^2n}{16\pi^2}G\tilde{G},
\end{equation}
where $g_X$ is the new gauge coupling, $n$ is the anomaly coefficient and $\tilde{G}$ is the dual of the field strength tensor $G$ of the new gauge group. This term breaks the shift symmetry of $\phi$ and generates a periodic potential of $\phi$ via instantons,
\begin{equation}
    V(\phi)=\Lambda^4\left(1-\cos\frac{\phi}{v_{B-L}}\right),
\end{equation}
The mass of the majoron can be found by,
\begin{equation}
    M_\phi^2=\frac{\partial^2 V}{\partial \phi^2}\Bigg|_{\phi=0}=\frac{\Lambda^4}{v_{B-L}^2}.
\end{equation}

We will assume that $M_\rho\sim v_{B-L}\gg T_{\rm RH}$ such that effects of $\rho$ can be ignored. We will also assume the majoron mass to always exceed twice the lightest RHN mass, $M_{\phi}>2M_{1}$. Moreover we consider a resonant leptogenesis scenario with $M_1\sim M_2\ll M_3$. The resonance effect makes sure that enough asymmetry is generated for all $M_1$ values starting from 100 GeV, to explain the BAU. For simplicity we take $M_3> M_\phi$. The decay width of $\phi$ to RH neutrinos is,
\begin{eqnarray}
\label{eq:phipartialdecaywidth}
\Gamma_{\phi\to N_i N_i} & = & \frac{M_{\phi}M_i^2}{16\pi v_{B-L}^2}\sqrt{1-\frac{4M_{i}^{2}}{M_{\phi}^{2}}}\,,
\label{eq:decaywidth_to_N}
\end{eqnarray}
where $i=1,2$ represents the 2 lightest RH neutrinos. The majoron decay to the hidden sector through the coupling in Eq.~\eqref{eq:majoron hidden coupling} is loop suppressed or forbidden if the hidden sector particles are heavier than $\phi$.

We consider the parameter space of the majoron where it dominates the energy budget of the Universe before decaying, leading to a phase of matter-dominated pre-BBN era. The RHNs produced from majoron decay further dominate the energy density. The decay of majoron and RHNs inject considerable amounts of entropy into the Universe. This dampens the primordial gravitational waves of inflatinary origin that re-enter the horizon during the early matter domination epochs. We show such signature of low-scale leptogenesis in primordial gravitational waves can be detected in various GW experiments such as BBO, DECIGO, $\mu$-ARES, ET, LISA, CE, etc.

\subsection{Set Up}
After inflationary reheating ends at temperature $T_{\rm RH}$, the majorons are produced either a) thermally via suppressed scatterings, e.g. $lH\rightarrow\phi N$ or b) non-thermally from inflaton and $\rho$-decay or through misalignment (coherrent oscillations) starting around $T_{\rm osc}\sim\sqrt{M_\phi M_{\rm Pl}}$. We make the coupling $y_N$ small such that $\phi$ is long-lived, hence no production through $N$ inverse decay is possible. The misalignment mechanism gives an energy density of the majoron,
\begin{equation}
    \rho_\phi(T_{\rm osc})\sim\frac{1}{2}M_\phi^2\phi_{\rm init}^2
\end{equation}
where $\phi_{\rm init}$ is the initial value of $\phi$ at $T_{\rm osc}$.
We consider that $\phi$ had an abundance equal to the thermal abundance at temperature $T\sim M_\phi$ after the end of inflation, i.e. $\rho_\phi(M_\phi)=M_\phi n_\phi^{eq.}$. We can adjust $\phi_{\rm init}$ to achieve this result. In thermal case, if scatterings freeze-out while the majoron particles were still relativistic we obtain the maximal $\phi$ abundance~\cite{Bezrukov:2009th, Kolb:1990vq}, 
\begin{equation}
    \label{eq:phiabundance}
    \frac{n_\phi}{s}\Big|_f = \frac{45 \,\zeta(3)}{2 \pi^4 \, g_{*S}}
\end{equation}
where subscript $f$ denotes the time of freeze-out, $n_\phi$ is the number density of $\phi$ and the entropy density is given by,
\begin{equation}
    \label{eq:entropydensity}
    s = \frac{2\pi^2}{45}g_{*S} T^3
\end{equation}
The effective number of degrees of freedom related to entropy is denoted by $g_{*S}$ which we take to be equal to $g_* =106.75$ at high temperatures for simplicity in our analysis. In both the cases: misalignment or thermal, $\phi$ is non-relativistic and frozen-out at $T\sim M_\phi$, after which its energy density red-shifts as matter and dominates over the radiation density starting from temperature~\cite{Giudice:1999fb,Kolb:1990vq}, 
\begin{equation}
    \label{eq:Tdom}
    T_{\rm dom} \sim \frac{M_\phi}{g_*} \approx 0.01 M_\phi
\end{equation}
The $\phi$-dominated phase starts at temperature $T_{\rm dom}$ and ends at decay temperature $T_\phi$, calculated by considering the sudden decay assumption,
\begin{equation}
  \label{eq:Tphi}
  T_{\phi} = \left(\frac{90}{8\pi^{3}g_{\ast}(T_{\phi})} \right)^{\frac{1}{4}}
  \sqrt{\Gamma_{\phi}M_\text{pl}}. 
\end{equation}
We always assume $T_\phi < T_{\rm dom}$ such that we get a period of majoron domination in the early Universe. The dynamics of $\phi$, RHNs and radiation can be obtained by solving the relevant Boltzmann equations which we will describe in Sec.~\ref{sec:boltz}. 

Now let us look at the aftermath of $\phi$ decay. We start with Eq.~\eqref{eq:phipartialdecaywidth}, the decay widths of $\phi$ to lightest RHNs. As long as the mass of the majoron $M_\phi>2M_{1,2}$, it can decay to $N_{1,2}$. The majorons we consider are essentially heavy ($10^5-10^{15}$ GeV) and non-relativistic in post-inflationary scenario and their decay is in their rest frame. Hence, RHNs resulting from this can reasonably be monochromatic, where each RHN will approximately have energy $E_N\sim M_\phi/2$ at production. We are concerned only about the decay dynamics of $N_1$~\cite{Engelhard:2007kf}. If $M_\phi\gg M_1$, the RHNs are relativistic when they are produced at temperatures around $T_\phi$. The energy of these relativistic RHNs, $E^{\rm rel}_N(T)$ redshifts with the expansion of the Universe such that $E^{\rm rel}_N(T)/T=E_N/T_\phi$. This gives,
\begin{equation}\label{eq:ENrel}
    E^{\rm rel}_N(z)\sim E_N\frac{T}{T_\phi}\sim\frac{M_\phi}{2\,T_\phi}\frac{M_1}{z},
\end{equation}
where $z=M_1/T$. The RHNs then become non-relativistic when their energy becomes equivalent to their mass, i.e. at temperature~\cite{Zhang:2023oyo},
\begin{equation}\label{eq:TNR}
    T_{\rm NR}\sim T_\phi\frac{2M_1}{M_\phi}.
\end{equation}
Hence, the energy of each RHN after its production can be written as,
  
\begin{equation}
    E_N(z)\sim
    \begin{cases}
        \frac{M_\phi}{2} & \text{for $T>T_\phi$} \\
        \frac{M_\phi}{2\,T_\phi}\frac{M_1}{z} & \text{for $T_\phi\geq T>T_{\rm NR}$} \\
        M_1 & \text{for $T\leq T_{\rm NR}$}
    \end{cases}
\end{equation} 
The RHNs further decay to lepton-Higgs pairs with decay widths~\cite{Cataldi:2024pgt},
\begin{equation}
  \label{eq:2}
   \Gamma_{N_1}=\Gamma_{N_1}^{rf}\, \frac{M_1}{E_N(z)}\sim
   \mathcal{H}(M_1)\, K \, \frac{M_1}{E_N(z)},
\end{equation}
where $\mathcal{H}(M_1)=\sqrt{\frac{8\pi^3g_\ast}{90}}\frac{M_1^2}{M_\text{pl}}$ is the Hubble parameter with $M_\text{pl}=1.22\times10^{19}$ GeV. Note that the decay rates get suppressed by the Lorentz boost or time dilation factor $\gamma_N=E_N(z)/M_1$.

The decay temperatures of the RHNs is calculated assuming that the energy density of RHN is converted instantaneously into radiation \cite{Hahn-Woernle:2008tsk},
\begin{equation}
\label{eq:TRHN}
   T_{N_1} = \left(\frac{90}{8\pi^{3}g_{\ast}(T_{N_1})} \right)^{\frac{1}{4}}
  \sqrt{\Gamma_{N_{1}}^{rf}M_\text{pl}} =M_1 \sqrt{K}
\end{equation}
where $g_\ast(T)$ is the relativistic degrees of freedom at the corresponding temperature $T$ which is taken to be $\sim106.75$. The final baryon asymmetry from RHN decay can be written as,
\begin{equation}
    \label{eq:YB}
    Y_B=\frac{n_B-n_{\overline{B}}}{s}=\frac{a_{\rm sph}\epsilon}{s}n_N
\end{equation}
where $s=g_*(2\pi^2/45)T^3$ is the entropy density and $n_B,n_{\overline{B}},n_N$ are the number densities of baryons, anti-baryons and RHNs respectively. $Y_B$ is related to $\eta_B$ given in Eq.~\eqref{eq:etaB} as, $Y_B = (n_{\gamma0}/s_0) \eta_B \simeq \eta_B/7.04$~\cite{Davidson:2008bu}, with $s_0$, $n_{\gamma0}$ being the current entropy and photon number densities respectively. The factor $a_{\rm sph} = 28/79$ is the efficiency of the sphaleron processes.

Using Eq.~\eqref{eq:phipartialdecaywidth} in Eq.~\eqref{eq:Tphi} and considering Eq.~\eqref{eq:Tdom}, we can write the condition $T_\phi<T_{\rm dom}$ in terms of couplings as,
\begin{equation}
    \label{eq:phi_dom_condition}
    y_{N_1}\lesssim0.2\sqrt{\frac{M_\phi}{M_{\rm pl}}}.
\end{equation}

In Table~\ref{tab:conversion}, we summarize all the necessary conversion formulas for the parameters used in our analysis in terms of the couplings and masses of $\phi$ and RHNs. We consider $M_\phi,T_\phi,M_1, K$ and $T_{\rm RH}$ as input parameters for our analysis and $v_{B-L}, y_{N_1},|\lambda|^2_{11},T_{N_1},\Gamma_\phi,\Gamma_{N_1}^{rf}$ are derived from them.
\begin{table}[!ht]
\begin{center}
\begin{tabular}{|c|c|c|}
	\hline
Variable & Symbol & Definition \\
	\hline
	\hline

RHN decay width in the rest frame& $\Gamma_{N_1}^{r f}$ & $\frac{|\lambda|^2_{11}M_1}{8\pi}$\\
Majoron decay width to RHNs & $\Gamma_{\phi\rightarrow N_iN_i}$ &  $\frac{y_{N_i}^2 M_{\phi}}{16\pi }\sqrt{1-\frac{4M_{i}^{2}}{M_{\phi}^{2}}}$ \\
Total decay width of majoron & $\Gamma_\phi$ & $\Gamma_{\phi\rightarrow N_1N_1}+\Gamma_{\phi\rightarrow N_2N_2}$ \\
Majoron decay temperature & $T_\phi$ & $\left(\frac{90}{8\pi^{3}g_{\ast}(T_{\phi})} \right)^{\frac{1}{4}}
  \sqrt{\Gamma_{\phi}M_\text{pl}}$ \\
  & & $\sim0.05\,y_{N_1}\left(\frac{100}{g_*(T_\phi)}\right)^{\frac{1}{4}}\sqrt{M_\phi M_{\rm pl}}$ \\
Washout parameter & $K$ & $\frac{\tilde{m}_1}{m_*}$=$\frac{\tilde{m}_1}{1.1\times10^{-3}\,\text{eV}}=\frac{|\lambda|^2_{11}v^2/M_1}{1.1\times10^{-3}\,\text{eV}}$\\
RHN decay temperature & $T_{N_1}$ & $\left(\frac{90}{8\pi^{3}g_{\ast}(T_{\phi})} \right)^{\frac{1}{4}}
  \sqrt{\Gamma_{N_1}^{r f}M_\text{pl}}\sim M_1\sqrt{K}$\\

			\hline \hline
\end{tabular}
\end{center}
\vspace{-0.17in}
\caption{\small \it Definition of the key parameters used for our analysis in terms of couplings and masses of the majoron and RHNs. Here $v=174$ GeV is the electroweak VEV, and $\lambda$ is the Yukawa coupling of RHNs with active neutrinos via SM Higgs. Here we also considered $y_{N_2}\sim y_{N_1}$.}
\label{tab:conversion}
\end{table}

\subsection{The Boltzmann Equations}\label{sec:boltz}
The relevant Boltzmann equations for the evolution of energy densities of the majoron, RHNs, radiation, and $B-L$ asymmetry are,
\begin{equation}
  \label{eq:boltzrho}
   \begin{aligned}
     \dot{\rho}_{\phi} &= -3\mathcal{H}\rho_{\phi}- \Gamma_{\phi}\,(\rho_{\phi}-\rho_{\phi}^{\rm eq}) \\
    \dot{\rho}_{N_1} &= -3\mathcal{H}\left(\rho_{N_1}+p_{N_1}\right)+\Gamma_{\phi\rightarrow N_1N_1}\,(\rho_{\phi}-\rho_{\phi}^{\rm eq})-\left(\Gamma_{N_1}+\Gamma_{\rm scatt}\right)(\rho_{N_1}-\rho_{N_1}^{eq}) \\
    \dot{n}_{B-L} &= -3\mathcal{H}n_{B-L}-\epsilon\, \Gamma_{N_1}(n_{N_1}-n_{N_1}^{eq})
    -\left(\Gamma_{ID}+\Gamma_{ID}^{\rm scatt}\right)\, n_{B-L} \\
    \dot\rho_{R} &= -4\mathcal{H}\rho_{R}+\left(\Gamma_{N_1}+\Gamma_{\rm scatt}\right)(\rho_{N_1}-\rho_{N_1}^{eq})\\
  \end{aligned}
\end{equation}
where $\mathcal{H}$ is the Hubble parameter in cosmic time and dot denotes the derivative with respect to cosmic time and $\rho^{eq}_X$ and $n_X^{eq}$ represents the equilibrium energy and number density of the species $X$ respectively. Here we consider the pressure of the RHN, $p_{N_1}=\rho_{N_1}/3$ for $T>T_{\rm NR}$ where $T_{\rm NR}$ is defined in Eq.~\eqref{eq:TNR}, and $p_{N_1}=0$ for $T\leq T_{\rm NR}$ , signifying the relativistic and non-relativistic evolution of the RHNs respectively. We will ignore the inverse decay term $\Gamma_\phi \rho_\phi^{\rm eq}$ since majoron decay temperature $T_\phi$ is much less than the majoron mass $M_\phi$, hence $\phi$ never comes into thermal equilibrium after it starts dominating. $\Gamma_{ID}$ is the inverse decay rate. See Appendix \ref{app:A} for details. 

In the above equations, we have written terms for decays, inverse decays and scattering processes of the non-thermally produced RHNs. Scatterings involving the majoron $lH\leftrightarrow N\phi$ are thermally suppressed at $T\ll M_\phi$. For $T_{\rm RH} \gg M_1$, thermal $N_1$ production via $\Delta L = 1$ scatterings (e.g.~$N_1 l \leftrightarrow q t$) is subdominant to non-thermal production from $\phi$ decay, since $\phi$ dominates the energy density prior to decay. Scatterings of non-thermal $N_1$ with the plasma, described by $\Gamma_{\rm scatt}=2\Gamma_{Ss}+4\Gamma_{St}$ and the associated washout $\Gamma_{ID}^{\rm scatt}=\Gamma_{Ss} \frac{\rho_{N_1}}{\rho_{N_1}^{eq}}+2\Gamma_{St}$~\cite{Buchmuller:2004nz,Zhang:2023oyo}, become relevant only when $\rho_{N_1}\sim \rho_R$, which happens at temperatures around $T_\phi$ and $T_{N_1}$. Here $\Gamma_{Ss}$ and $\Gamma_{St}$ are the $s-$ and $t-$shannel scattering rates involving top quark and gauge bosons. We solved the above Boltzmann equations with and without the scattering terms and found that, within the parameter space of interest, the inclusion of these effects modifies the final baryon asymmetry mostly by $<10\%$. Only at $K\sim20$, the change is $\sim50\%$. This means $N_1$ decays and inverse decays dominate the dynamics compared to scatterings. A full kinetic treatment (see e.g.~\cite{Hahn-Woernle:2009jyb, Baker:2024udu}) is beyond the scope of this paper, hence we neglect scatterings in the rest of the analysis.

We solve the Boltzmann equations from the time when the majoron energy density overtakes the radiation energy density. Thus the initial energy density of the majoron is equal to the radiation density~\cite{Kolb:1990vq},
\begin{equation}
    \rho_{\phi}(T_{\rm dom})=\rho_{R}(T_{\rm dom})=\frac{\pi^2}{30}g_{\ast}(T_{\rm dom})\,T_{\rm dom}^4
\end{equation}
We take $T_{\rm dom}$ from Eq.~\eqref{eq:Tdom}. The initial density of RHN is taken to be free parameter smaller than $\rho_{\phi}(T_{\rm dom})$ while initial $B-L$ asymmetry is taken to be negligible. Further details on the numerical solution of the Boltzmann equations and calculation of the final $\kappa_f$ value are given in Appendix \ref{app:A}. In Fig.~\ref{fig:kappaThermal}, along with the thermal case, we show the enhancement in $\kappa_f$ in the non-thermal leptogenesis scenario by solving the Boltzmann equations. The solid magenta line and dashed green line show the enhancement of the final efficiency factor as a function of the washout parameter for $M_1=10^5$ GeV, $M_\phi=10^7$ GeV and $T_\phi=10^5,3\times10^4$ GeV in the non-thermal scenario. One can compare with thermal leptogenesis scenario (solid red and dashed blue lines). We notice that the value of washout parameter $K$ at which $\kappa_f$ is maximum increases with decreasing $T_\phi$ for the given parameters in Fig.~\ref{fig:kappaThermal}, in contrast to results found in~\cite{Hahn-Woernle:2008tsk,Zhang:2023oyo}. This is because, unlike~\cite{Hahn-Woernle:2008tsk,Zhang:2023oyo}, we have considered the relativistic phase of RHNs in the Boltzmann equations and our RHNs decay width given in Eq.~\eqref{eq:2} is not thermally averaged.

At this point, we can numerically calculate the final baryon asymmetry in our model. However, we are not interested in the exact amount of asymmetry generated, since our goal is studying the GW probe of leptogenesis. In the next section, we discuss the classification of scenarios based on numerical results and analytical approximations, relevant to GW sensitivity.

\begin{figure}[!ht]
\centering
\includegraphics[width=0.7\linewidth]{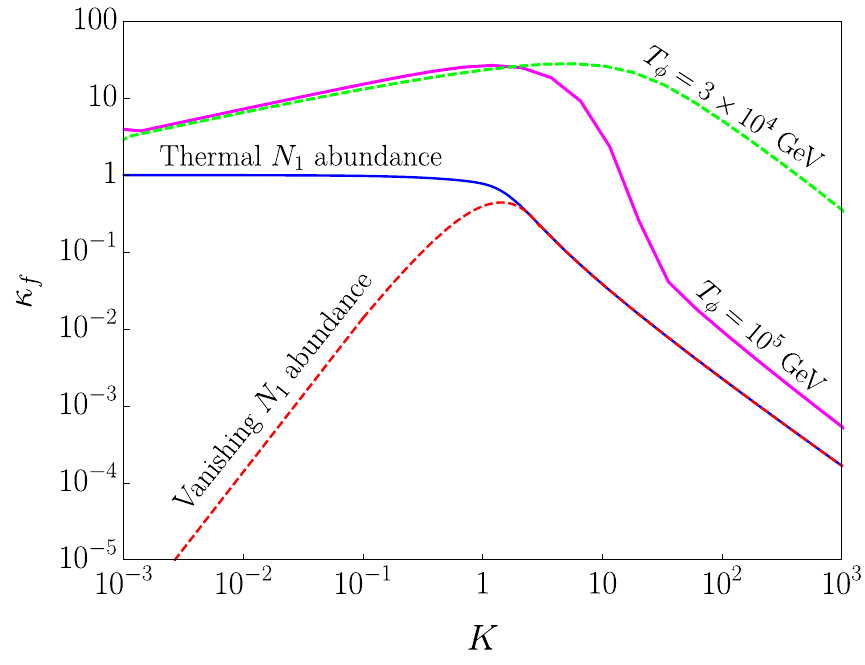}  
\caption{\small \it Final efficiency factor $\kappa_f$ as a function of washout parameter $K$ in thermal and non-thermal leptogenesis. The solid blue and the dashed red curves represent a thermal and vanishing initial abundance of $N_1$ respectively in thermal leptogenesis. Here we used Eq.~\eqref{eq:kappaf_thermal} and~\eqref{eq:zB} for the blue curve while for the red curve, we use Eq.~\eqref{eq:kappaf_weak_vanishing} for $K\ll1$ and interpolate it to the thermal case in the $K\gg1$ regime. The solid magenta and dashed green lines are solutions to the Bolzmann equations~\eqref{eq:boltzrho} for non-thermal leptogenesis with $M_1=10^5$ GeV, $M_\phi=10^7$ GeV for $T_\phi=10^5$ GeV, and $3\times10^4$ GeV respectively.}
    \label{fig:kappaThermal}
\end{figure}


\section{Classification of Scenarios}\label{sec:classification}
In this section, we identify three general non-thermal scenarios relevant for GW observations along the lines of~\cite{Zhang:2023oyo}. We present analytical estimates of the final asymmetry, equilibrium temperature when matter domination ends, and the dilution factor from entropy injection for each of the four scenarios.
In brief, the four scenarios are,

\begin{enumerate}
    \item \textit{Case (a): Instantaneous RHN decay -} Portrayed in Fig.~\ref{fig:densityCaseA}, we classify this scenario by requiring $\Gamma_N \gg \Gamma_\phi$ which implies $T_{\rm N_1} \sqrt{\frac{2 M_1}{M_\phi}} \gg T_\phi$ using Eq.~\ref{eq:2}. This means that the RHNs are mostly produced at temperatures around $T_\phi$ when the majoron completely decays away and radiation starts dominating again. Those RHNs instantaneously decay to the SM bath because the lifetime of the highly relativistic RHNs, even after time dilation by Lorentz factor $\gamma_N = M_\phi/2M_1$, is still shorter than the lifetime of majoron $\phi$. In the non-relativistic RHN limit, i.e. $\gamma_N \rightarrow 1$, the condition for instantaneous RHN decay becomes $T_{N_1}\gg T_\phi$. In both relativistic and non-relativistic scenarios, RHNs do not have time to thermalize with the radiation bath even in the strong washout regime $K\gg1$, thus the RHN decay process is strongly out-of-equilibrium. The evolution of the energy densities of different matter and radiation are shown in Fig. \ref{fig:densityCaseA} for both weak ($K\ll1$) and strong ($K\gg1$) washout regimes. We can see that for a strong washout, the RHN density is lower than the weak washout case due to a higher decay rate of the RHNs. The final asymmetry in both the weak and strong washout regions can be estimated analytically by finding the value of $n_N$ at temperature $T_\phi$ by equating $\rho_\phi=\rho_R$ and $n_N = 2\, n_\phi$ as~\cite{Zhang:2023oyo},
    \begin{equation}\label{eq:YBcaseA}
        Y_B \equiv \frac{n_B - n_{\overline{B}}}{s} = \frac{3}{2} a_{\rm sph} \epsilon \frac{T_\phi}{M_\phi}
    \end{equation}
    In this case, one can assume entropy is injected directly into the SM thermal bath at the time of majoron decay since RHN decay is instantaneous.

    \begin{figure}[!ht]
        \centering
            \begin{subfigure}{0.5\linewidth}
        \includegraphics[width=\linewidth]{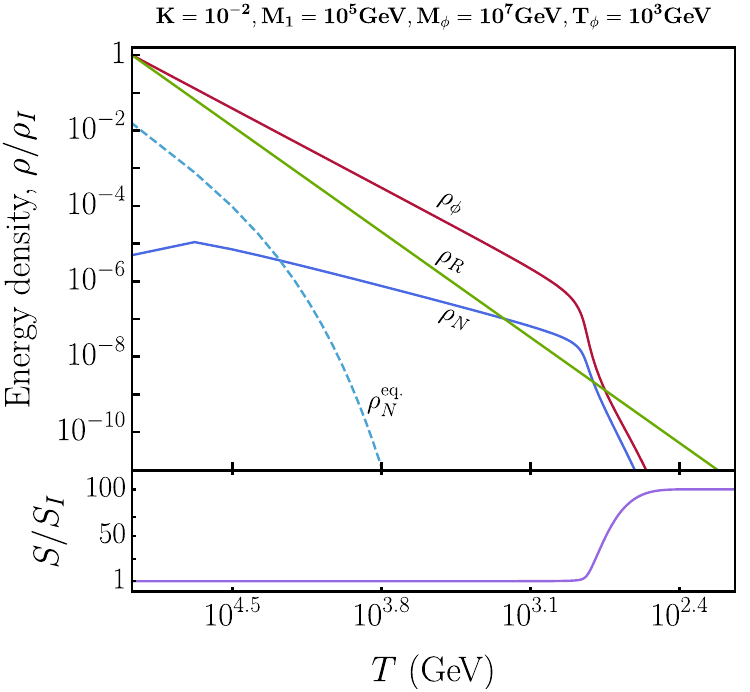} 
            \label{fig:CaseAK<1}
            \end{subfigure}\hfill
            \begin{subfigure}{0.5\linewidth}
        \includegraphics[width=\linewidth]{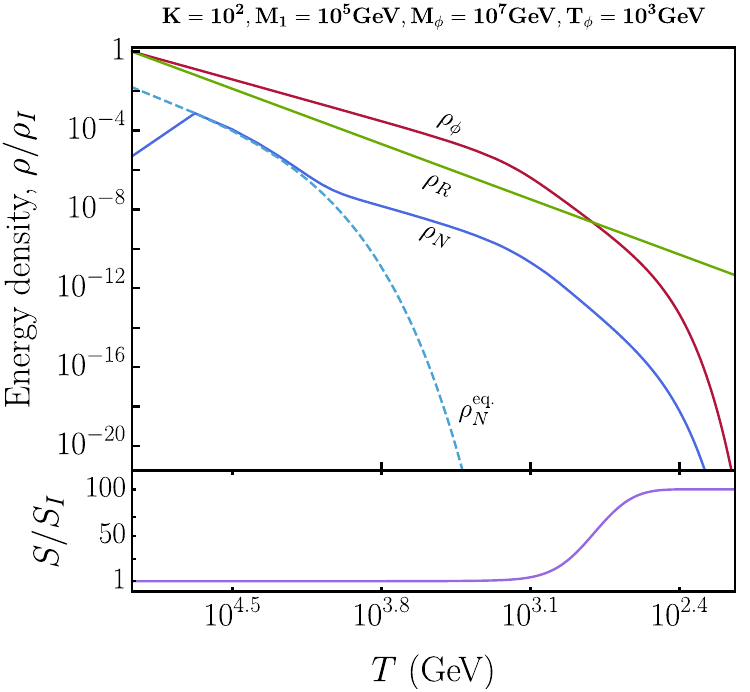}
            \label{fig:CaseAK>1}
            \end{subfigure}\hfill
            
        \caption{\small \it The evolution of the energy densities of the majoron $\rho_\phi$, RHN $\rho_N=\rho_{N_1}$ and thermal radiation, $\rho_R$ with respect to the thermal radiation temperature. Here we have taken $M_1=10^5, M_\phi=10^7,T_{\phi}=10^{3}$ GeV  for weak washout $K=10^{-2}$ (\textbf{Left}) and strong washout $K=10^2$ (\textbf{Right}). Both correspond to Case (a) mentioned at the beginning of Sec. \ref{sec:classification}. We can see RHNs instantaneously decay after production. The bottom panels in each of the images show the total entropy of the Universe, $S=s_{\rm total}a^3$ where $s_{\rm total}=s+s_{\rm N}$, with $s$ corresponding to the thermal sector entropy density, defined in Eq.~\eqref{eq:entropydensity} and $s_N$ being the entropy density of the decoupled RHN sector. Here $a$ is the scale factor.}
            \label{fig:densityCaseA}
        \end{figure}

    \item \textit{Case (b): RHN Radiation Domination -} For this scenario we need $\Gamma_N \ll \Gamma_\phi$ implying $T_{N_1}\sqrt{\frac{2 M_1}{M_\phi}} \ll T_\phi$ which means the RHN decay is delayed after production. The RHNs dominate the energy budget as radiation after the majoron $\phi$ decays. The RHNs will decay relativistically at temperature $T_N^{\rm rel}$ when their decay rate becomes equal to the Hubble rate, i.e. $\Gamma_N(T_N^{\rm rel})\sim\mathcal{H}(T_N^{\rm rel})$. Considering $\mathcal{H}(T_N^{\rm rel})\sim (T_N^{\rm rel})^2/M_{\rm pl}$ and using Eq.s~\eqref{eq:ENrel},~\eqref{eq:2} and~\eqref{eq:TRHN}, we obtain the decay temperature of RHNs to be,
    \begin{equation}\label{eq:TNrel}
        T_{N}^{\rm rel}\sim \left(T_{N_1}^2\,T_{\rm NR}\right)^{1/3}
    \end{equation}
    The necessary condition for RHN radiation domination is $T_{\rm NR}<T_{N_1}$ which also implies $T_N^{\rm rel}<T_{N_1}$ as shown in Fig.~\ref{fig:densityCaseB}. The (\textbf{Left}) plot shows the RHN radiation domination for $K\ll1$ whereas the (\textbf{Right}) shows RHN radiation domination for $K\gg1$.
    In the weak washout regime ($K\ll1$), one can  calculate $n_N$ at $T_N^{\rm rel}$ by considering $\rho_N =\rho_R$, where $\rho_N$  can be written as $E_N^{\rm rel} n_N$. Using Eq.~\eqref{eq:YB}, we find that the final asymmetry turns out to be the same as Case (a) above, given in Eq.~\eqref{eq:YBcaseA}. For $K\gg 1$, the final asymmetry can be calculated by solving the Boltzmann equations. Since the RHNs decay relativistically, there is negligible entropy injection at $T_N^{\rm rel}$. However, significant entropy is injected into the RHN sector from the majoron decay at around temperature $T_\phi$.

    \begin{figure}[!ht]
        \centering
            \begin{subfigure}{0.5\linewidth}
        \includegraphics[width=\linewidth]{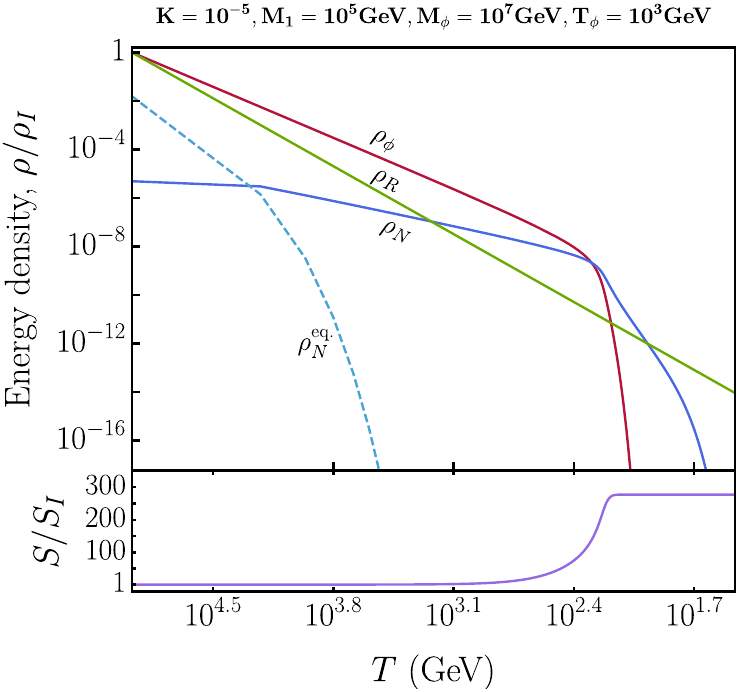} 
            \label{fig:CaseBK<1}
            \end{subfigure}\hfill
            \begin{subfigure}{0.5\linewidth}
        \includegraphics[width=\linewidth]{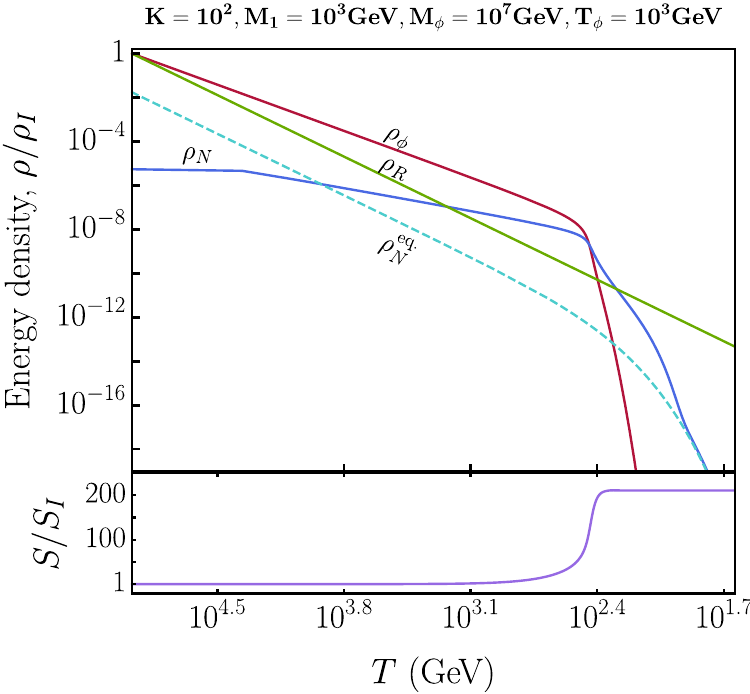}
            \label{fig:CaseBK>1}
            \end{subfigure}\hfill
            
        \caption{\small \it{Same as Fig. \ref{fig:densityCaseA} but for RHN radiation domination scenario, as described in Case (b) of Sec. \ref{sec:classification}, with $M_\phi=10^7$ and $T_\phi=10^3$ GeV assuming $M_1=10^5$ GeV, $K=10^{-5}$ (\textbf{Left}) and $M_1=10^3$ GeV, $K=100$ (\textbf{Right}). RHNs dominate as radiation starting from temperature $T_\phi$ till their decay at $T_N^{\rm rel}$ defined in Eq.~\eqref{eq:TNrel}. For weak washout, $K\ll1$ (\textbf{Left}), RHNs never themalizes however the figure on (\textbf{Right}) shows RHN radiation domination for $K\gg1$, where RHNs can thermalize later due to strong Yukawa couplings. The bottom panels show the entropy injection.}}
            \label{fig:densityCaseB}
        \end{figure}

    \item \textit{Case (c): RHN Matter Domination -} We again consider $\Gamma_N\ll\Gamma_\phi$ for this scenario as Case (b), i.e. $T_{N_1}\sqrt{\frac{2M_1}{M_\phi}}\ll T_\phi$. However, unlike Case (b), here we require $T_{N_1}\ll T_\phi$, more specifically we are interested in the scenario where $T_{N_1}<T_{\rm NR} < T_\phi$ which translates to $\sqrt{K} < 2 T_\phi/M_\phi$. Thus strong washout $K>1$ is not possible in this scenario in our model. Relativistic RHNs first dominate the energy budget of the Universe as radiation after the decay of the majoron $\phi$. Then the RHNs become non-relativistic at $T_{\rm NR}$ and dominate as matter until their decay at temperature $T_{N_1}$. Hence there are two matter-dominated epochs separated by a radiation-dominated epoch and hence two stages of entropy injection in this case, first when the majoron decays to relativistic RHNs at $T_\phi$ and second when the RHNs decay at $T_{N_1}$. In the first stage, entropy is injected into the decoupled RHN sector and not into the SM thermal bath. When the RHNs decay out-of-equilibrium, the final radiation domination starts. If $K \ll 1$ the RHNs never thermalize with the radiation bath as seen in Fig. \ref{fig:densityCaseC} (\textbf{Left}). The final BAU for $K \ll 1$ is estimated as~\cite{Zhang:2023oyo},
    \begin{equation}\label{eq:YBcaseC}
        Y_B = \frac{3}{4} a_{\rm sph} \epsilon \frac{T_{N_1}}{M_1} \simeq 0.26 \sqrt{K}\epsilon
    \end{equation}
    In Fig.~\ref{fig:leptoAsym}, we show the parameter space of the lightest RHN, spanned by $M_1$ and $K$ and the regions of successful non-thermal leptogenesis within it, i.e. $\eta_B\geq\eta_B^{\rm CMB}$, for different values of $M_\phi$ and $T_\phi$. The darker gray region at the bottom left is where sphaleron processes are turned off and leptogenesis is not efficient. The different colored regions correspond to different values of $M_\phi$ and $T_\phi$. For this plot we utilize analytical expressions given in Eq.~\eqref{eq:YBcaseA},~\eqref{eq:YBcaseC} for Case (a), Case (c), and Case (b) $K<1$ regions, and numerical solutions of the Boltzmann equations for Case (b) $K>1$ region.

    \begin{figure}[!ht]
        \centering
        \includegraphics[width=0.7\linewidth]{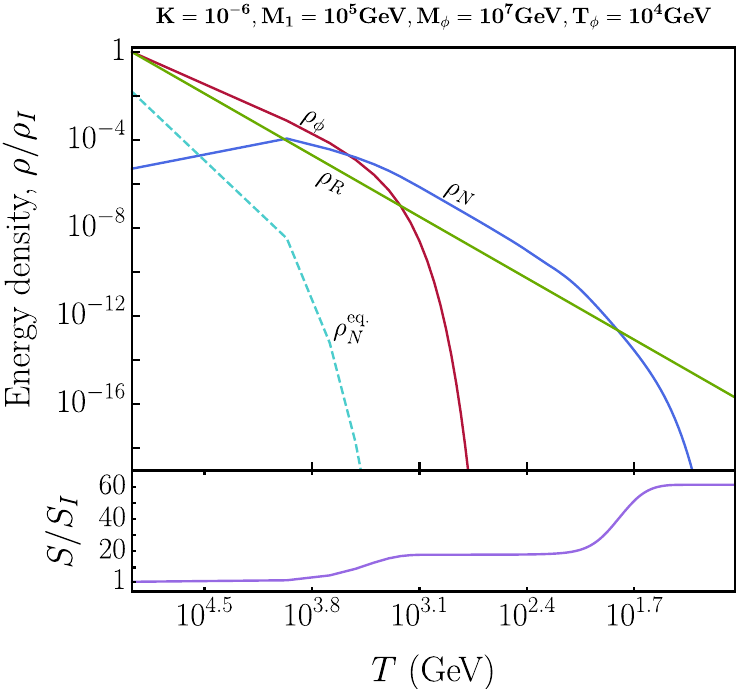} 
        \caption{\small \it{Same as Fig. \ref{fig:densityCaseA} but for RHN matter domination scenario, described in Case (c) of Sec. \ref{sec:classification}, with $K=10^{-6}$, $M_1=10^5$ GeV, $M_\phi=10^7$ GeV and $T_\phi=10^4$  GeV. RHNs dominate as radiation first then as matter after temperature $T_{\rm NR}\sim 200$ GeV, defined in Eq.~\eqref{eq:TNR}. The bottom panels show the two stages of entropy injection.}}
            \label{fig:densityCaseC}
        \end{figure}

    \begin{figure}[!ht]
        \centering
        \includegraphics[width=0.7\linewidth]{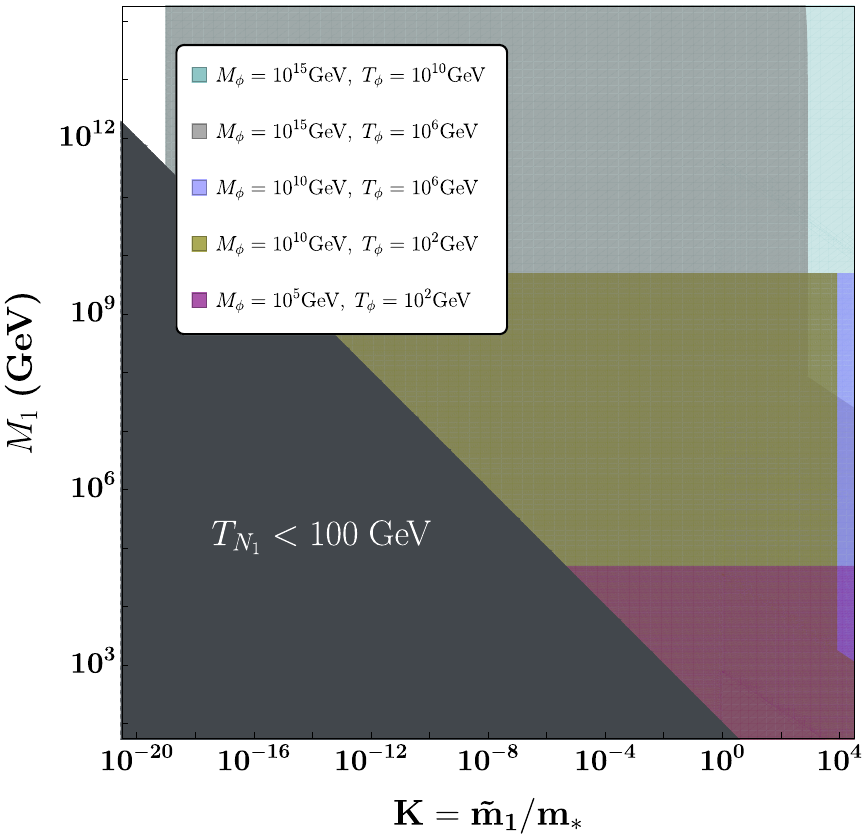} 
        \caption{\small \it{Shaded regions of different colors correspond to successful non-thermal leptogenesis, $\eta_B\gtrsim\eta_B^{\rm CMB}$, in the parameter space of the lightest RHN for given values of $M_\phi$ and $T_\phi$. The dark gray region at the left bottom is where the RHN decay temperature is less than 100 GeV, hence sphaleron processes are suppressed and leptogenesis is not efficient.}}
            \label{fig:leptoAsym}
        \end{figure}
\end{enumerate}

\begin{figure}[!ht]
        \centering
        \includegraphics[width=\linewidth]{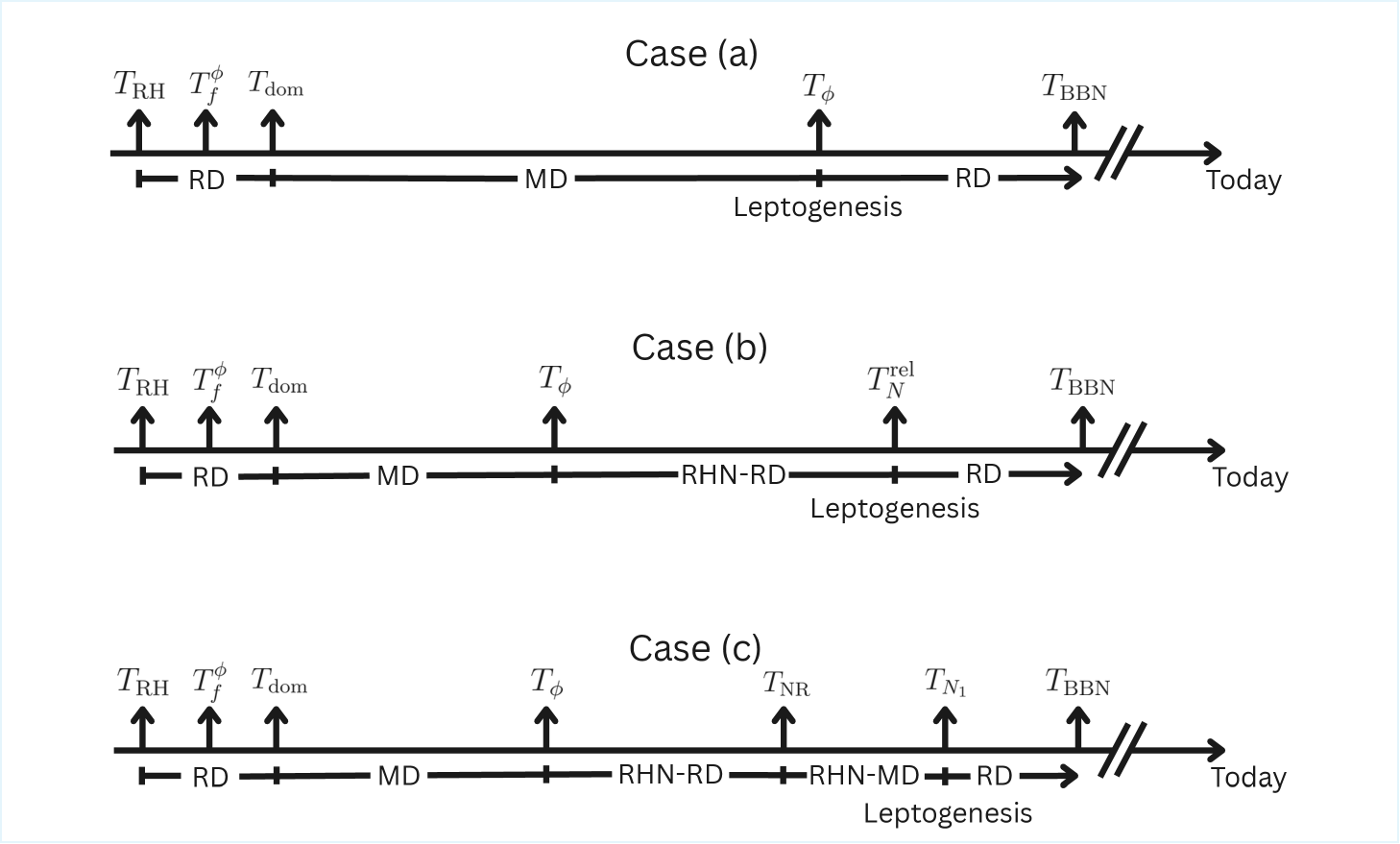} 
        \caption{\small \it{Schematics of key events and temperatures for Case (a), (b) and (c). Inflation ends at reheating temperature $T_{\rm RH}$ after which $\phi$ remains in thermal equilibrium till its freeze out temperature $T_f^\phi$. Then $\phi$ becomes non-relativistic and start the first matter-domination epoch at $T_{\rm dom}\sim1\%M_\phi$}. This matter-domination ends when $\phi$ decays at $T_\phi$. For case (a), RHNs also instantaneuously decay at $T_\phi$ and leptogenesis occurs before the BBN temperature $T_{\rm BBN}$. In case (b), RHNs dominate as radiation from $T_\phi$ till their decay at $T_N^{\rm rel}$. In case (c), RHNs dominate as radiation till $T_{\rm NR}$ at which point they become non-relativistic and start the second MD epoch. This epoch ends at RHN decay temperature $T_{N_1}$ when leptogenesis occurs.}
            \label{fig:schematic}
        \end{figure}

\subsection{Analytical Estimates of Equilibrium Temperature}
Equilibrium temperature is the temperature at which a matter-dominated epoch ends, kickstarting a radiation-dominated era. From the discussion above, we see that for Case (a) and (b), the equilibrium temperature is around the decay temperatures $T_\phi$. The exact equilibrium temperature is a little lower than the decay temperature since the decay is not instantaneous. However, for our considerations, $T_\phi$ is a reasonable estimate of the equilibrium temperature.

In Cases (c), there are two matter-dominated epochs, hence entropy is injected into the Universe in two steps as discussed earlier. First, the majoron decaying to relativistic RHNs increases the total entropy of the Universe at around $T_\phi$. The second stage of entropy injection is when RHNs decay at temperature $T_{N_1}$. There is a radiation-dominated epoch between the two matter-dominated epochs. Hence we take the two equilibrium temperatures to be $T_\phi$ and $T_{N_1}$. Thus we approximate the equilibrium temperatures for GW analysis as,
\begin{equation}
    \label{eq:Teq}
    T_{\rm eq.}\equiv T_\text{dec.} \sim \begin{cases}
        T_\phi\quad\text{for Case (a) and (b)}\\
        T_\phi\,\&\, T_{N_1}\quad\text{for Case (c)}
    \end{cases}   
\end{equation}

\subsection{Dilution Factor from Entropy Injection}
Both majoron and non-relativistic RHN decay inject entropy into the Universe if the decay products are relativistic. The general formula for the dimensionless dilution factor from the entropy injection by a dominating species $\chi$ decaying to relativistic particles is given by ~\cite{Scherrer:1984fd,Berbig:2023yyy}
\begin{equation}
\label{eq:dilution_factor}
    \Delta = \frac{s(T_\text{after})a^3(T_\text{after})}{s(T_\text{before})a^3(T_\text{before})}=\left(1+2.95\left(\frac{2\pi^2 \left<g_*(T)\right>}{45}\right)^{\frac{1}{3}}\frac{(\frac{n_\chi}{s}M_\chi)^{\frac{4}{3}}}{(M_\text{pl}\Gamma_\chi)^{\frac{2}{3}}}\right)^{\frac{3}{4}}
\end{equation}
where $s$ is the entropy density, $T_\text{before/after}$ represents temperature before/after the decay process and $\frac{n_\chi}{s}$ is the initial abundance of the particle $\chi$. 

For our analysis, we estimate the dilution factor using the formula in~\eqref{eq:dilution_factor} for the four classifications we made: 

\textit{Case (a):} In this case, the RHNs decay out-of-equilibrium instantaneously to relativistic particles at $T_\phi$. Therefore we can assume that it is the majoron that is directly decaying to the radiation bath releasing entropy. The dilution factor $\Delta$ can be given by Eq.~\eqref{eq:dilution_factor} with $\chi$ replaced by $\phi$. The initial abundance is given by Eq. \eqref{eq:phiabundance}. In the large entropy injection approximation $\Delta\gg1$, using Eq.~\eqref{eq:Tphi} we can write the dilution factor to be,
\begin{equation}
    \label{eq:CaseA_Delta}
    \Delta \simeq 3.7\times10^9 \left(\frac{M_\phi}{10^{15}\text{ GeV}}\right)\left(\frac{ \text{TeV}}{T_\phi}\right)
\end{equation}

\hspace{1px}

\textit{Case (b):} RHNs dominate the energy budget of the Universe as radiation after their production. Eventually, the RHNs decay relativistically at $T_{N}^{\rm rel}<T_\phi$. Hence the RHN decay injects negligible entropy into the Universe. The only stage of substantial entropy injection is when the majoron $\phi$ decays at $T_\phi$. The dilution factor is again estimated by Eq.~\eqref{eq:CaseA_Delta}.

\hspace{1px}

\textit{Case (c):} RHNs dominate the energy budget of the Universe as soon as they are produced, first as radiation, then as matter. The energy density stored in $\phi$ species is first transferred to the RHNs and then dumped into the radiation bath when RHNs decay at $T_{N_1}$. Hence, this scenario leads to two matter-dominated epochs separated by a radiation-dominated epoch.

The dilution factor from $\phi$ decay is again given by Eq.~\eqref{eq:CaseA_Delta}. The dilution factor from RHN decay can be estimated using Eq.~\eqref{eq:dilution_factor} considering the RHN decay width $\Gamma_N^{rf}$ and initial abundance after $\phi$ decay as,
\begin{equation}
    \frac{n_N}{s_{\rm tot}^{\text{after}}}=\frac{2 \, n_\phi}{\Delta_\phi \, s_{\rm tot}^{\rm before}}=\frac{2}{\Delta_\phi}\frac{n_\phi}{s}=\frac{45}{\pi^4}\frac{\zeta(3)}{g_{*S}}\frac{1}{\Delta_\phi}
\end{equation}
where $s_{\rm tot}$ is the total entropy of the Universe including entropy of the decoupled RHN sector, and superscripts ``before'' and ``after'' refers to before and after $\phi$ decay. Total entropy after $\phi$ decay, $s_{\rm tot}^{\rm after}$ can be written as $s_{\rm tot}^{\rm before}\Delta_\phi$ where $\Delta_\phi$ is given by Eq.~\eqref{eq:CaseA_Delta}. For the last equality, we use Eq.~\eqref{eq:phiabundance} since before the majoron decays the total entropy of the Universe is simply the entropy of the thermal bath. Note that considering only the lightest RHN $N_1$ is sufficient here. Again in the large entropy injection limit $\Delta_N\gg1$, one can write the dilution factor from RHN decay as,
\begin{equation}
    \label{eq:CaseC_Delta}
    \Delta_N \simeq 2000\,\left(\frac{10^{-12}}{K}\right)^{1/2}\left(\frac{10^8\text{ GeV}}{M_\phi}\right)\left(\frac{T_\phi}{100\text{ TeV}}\right)
\end{equation}


\section{Gravitational Waves of Inflationary Origin}\label{sec:distortion}
\subsection{Primordial spectrum in standard and non-standard cosmology}

Primordial inflation, the early exponential expansion of the Universe, is believed to have transitioned into an epoch known as reheating. This transition resulted in the creation of a high-energy plasma of radiation, following the dynamics of the Standard Model of particle physics. During inflation, gravitational waves (GWs) from quantum tensor fluctuations of the metric were generated. Initially, while outside the cosmological horizon, these GWs exhibited constant amplitudes. However, as they re-entered the horizon in the radiation-dominated era, their amplitudes underwent damping. The propagation of GW modes that re-enter the horizon is controlled by a transfer function.


The current power spectrum of GWs, denoted as $\Omega_{GW}(k)$ can be expressed as a function of the wave number, $k=2\pi f$ where $f$ is the frequency \cite{Berbig:2023yyy},

\begin{equation}\label{eq:OmegaGW}
    \Omega_{GW}(k) = \frac{1}{12} \left(\frac{k}{a_0 \mathcal{H}_0}\right)^2 T_T^2(k) P_T^\text{prim.}(k)
\end{equation}
with $a_0=1$ representing the present-day scale factor and $\mathcal{H}_0 \approx 2.2 \times 10^{-4}$ Mpc$^{-1}$ being the current Hubble expansion rate \cite{Datta:2022tab}. $P_T^\text{prim.}(k)$ is the primordial tensor power spectrum from inflation, and $T_T^2(k)$ is the transfer function that characterizes the propagation of GWs in the background of a Friedmann-Lemaitre-Robertson-Walker (FLRW) Universe after the temperature at the time of horizon re-entry, $T_\text{in}$ given by \cite{Nakayama:2008wy}:

\begin{equation}
    T_\text{in} = 5.8 \times 10^6 \, \text{GeV} \left(\frac{106.75}{g_*(T_\text{in})}\right)^{1/6} \left(\frac{k}{10^{14} \, \text{Mpc}^{-1}}\right)
\end{equation}

The conventional parameterization for the inflationary tensor power spectrum, $P_T^\text{prim.}(k)$ typically involves expressing its amplitude ($A_T$) and spectral index ($n_T$) with respect to the pivot scale $k_* = 0.05 \, \text{Mpc}^{-1}$, as outlined in the Planck 2018 analysis \cite{Planck:2018jri}.

\begin{equation}\label{eq:PTprime}
    P_T^\text{prim.}(k) = A_T(k_*) \left(\frac{k}{k_*}\right)^{n_T}
\end{equation}
where the amplitude $A_T(k_*)$ is related to the scalar power spectrum via the tensor-to-scalar-ratio $r \leq 0.035$ \cite{BICEP:2021xfz}:

\begin{equation}
    A_T(k_*) = A_S(k_*) \, r, \quad A_S(k_*)=2.0989 \times 10^{-9}
\end{equation}
where $A_S(k_*)$ is scalar power spectrum \cite{Planck:2018jri}. For our analysis, we fix the value of $r=0.035$. The spectral index $n_T$ plays a critical role in characterizing the inflationary tensor power spectrum. While standard single-field slow-roll inflation predicts a red-tilted spectrum, where $n_T$ satisfies the consistency relation $n_T = -r/8$ \cite{Liddle:1993fq}, other scenarios, including blue-tilted spectra ($n_T > 0$), are well-motivated in various cosmological models \cite{Brandenberger:2006xi, Baldi:2005gk, Kobayashi:2010cm, Calcagni:2004as, Calcagni:2013lya, Cook:2011hg, Mukohyama:2014gba, Kuroyanagi:2020sfw}.

Numerical computation and analytical fitting give the transfer function $T_T^2(k)$ as:

\begin{equation}
    T_T^2(k) = \Omega_m^2 \left(\frac{g_*(T_\text{in})}{g_*^0}\right) \left(\frac{g_{*S}^0}{g_{*S}(T_\text{in})}\right)^{4/3} \left(\frac{3j_1(z_k)}{z_k}\right)^2 F(k)
\end{equation}
Here, $g_*^0 = 3.36$ and $g_{*S}^0 = 3.91$ are the present-day values of the effective degrees of freedom. $\Omega_m = 0.31$ \cite{Datta:2022tab} is the total matter density. The Bessel function $j_1(z_k)$ describes the damping of GW amplitudes after horizon re-entry with $z_k\equiv k\tau_0$ and $\tau_0=2/\mathcal{H}_0$. In the limit $z_k \gg 1$, relevant for the frequencies of interest, we can replace $j_1(z_k)$ for $1/(\sqrt{2}z_k)$. Finally, the fitting function $F(k)$ in standard cosmology is given by \cite{Kuroyanagi:2014nba},
\begin{equation}
    F(k)_\text{standard} = T_1^2\left(\frac{k}{k_\text{eq.}}\right)T_2^2\left(\frac{k}{k_\text{RH}}\right)
\end{equation}
If there was an intermediate matter domination epoch, the fitting function is modified \cite{Kuroyanagi:2014nba},
\begin{equation}\label{eq:FIMD}
F(k)_\text{IMD} = T_1^2\left(\frac{k}{k_\text{eq.}}\right)T_2^2\left(\frac{k}{k_\text{dec.}}\right)T_3^2\left(\frac{k}{k_\text{dec. S}}\right)T_2^2\left(\frac{k}{k_\text{RH S}}\right)
\end{equation}
Here we introduce:

\begin{align}\label{eq:gwkeq}
    k_\text{eq.} &= 7.1 \times 10^{-2} \, \text{Mpc}^{-1} \cdot \Omega_m h^2 \\
    \label{eq:gwkdec}
    k_\text{dec.} &= 1.7 \times 10^{14} \, \text{Mpc}^{-1} \left(\frac{g_{*S}(T_\text{dec.})}{g_{*S}^0}\right)^{1/6} \left(\frac{T_\text{dec.}}{10^7 \, \text{GeV}}\right) \\
    \label{eq:gwkRH}
    k_\text{RH} &= 1.7 \times 10^{14} \, \text{Mpc}^{-1} \left(\frac{g_{*S}(T_\text{RH})}{g_{*S}^0}\right)^{1/6} \left(\frac{T_\text{RH}}{10^7 \, \text{GeV}}\right) \\
    \label{eq:gwkdecS}
    k_\text{dec. S} &= k_\text{dec.} \Delta^{2/3} \\
    \label{eq:gwkRHS}
    k_\text{RH S} &= k_\text{RH} \Delta^{-1/3}
\end{align}
where $T_\text{dec.}$ is defined by \eqref{eq:Teq} and we set $h = 0.7$. The entropy dilution factor $\Delta$ is the dilution factor and the fit functions read,
\begin{align}
    T_1^2(x) &= 1 + 1.57x + 3.42x^2 \\
    T_2^2(x) &= \left(1 - 0.22x^{3/2} + 0.65x^2\right)^{-1} \\
    T_3^2(x) &= 1 + 0.59x + 0.65x^2
\end{align}

At this point we can plot the spectrum $\Omega_{\rm GW}h^2$ as a function of current frequency, using Eq.~\eqref{eq:OmegaGW}. We discuss the features in these plots accounting for $\phi$ and $N$ domination in Sec.~\ref{sec:resultsSpec}. The suppression of the GW spectrum happens above a frequency $f_{\rm sup}$, given by,
\begin{equation}
    \label{eq:fsup}
    f_\text{sup}\simeq 2.7\times10^{-8}\;\text{Hz}\; \left(\frac{T_\text{dec.}}{{\rm GeV}}\right)
\end{equation}
The suppression factor is given by~\cite{Seto:2003kc},
\begin{equation}
    \label{eq:Rsup}
    R_\text{sup} = \frac{\Omega_\text{GW}^\text{IMD}}{\Omega_\text{GW}^\text{standard}}\simeq \frac{1}{ \Delta^{4/3}}
\end{equation}


\subsection{Two-step entropy injection}
Our Case (c) discussed in Sec. \ref{sec:classification} have two stages of entropy injection. We could not find any previous study of the transfer function to fit two stages of entropy injection. Therefore, extrapolating on results obtained from numerical fitting in~\cite{Kuroyanagi:2014nba}, we propose a modified transfer function to incorporate the two stages by assuming the form of the fitting function analogous to Eq.~\eqref{eq:FIMD} to be,
\begin{equation}\label{eq:2stepF}    
F(k)_\text{IMD}^\text{2-step} = T_1^2\left(\frac{k}{k_\text{eq.}}\right)T_2^2\left(\frac{k}{k_\text{dec.}^\phi}\right)T_3^2\left(\frac{k}{k_\text{dec.S}^\phi}\right) T_2^2\left(\frac{k}{k_\text{dec.}^N}\right)T_3^2\left(\frac{k}{k_\text{dec.S}^N}\right) T_2^2\left(\frac{k}{k_\text{RH S}^\text{2-step}}\right)
\end{equation}
where $k_\text{dec.}^{\phi},k_\text{dec.}^{N}$ are related to the decay temperatures $T_\phi$ and $T_{N_1}$ respectively and $k_\text{dec.S}^\phi,k_\text{dec.S}^{N}$ are related to the domination temperatures of $\phi$ and $N_1$ respectively. Note that in Eq.~\eqref{eq:2stepF}, we use the function $T^2_2(x)$ for the energy scales corresponding to decay temperatures and $T^2_3(x)$ for energy scales corresponding to domination temperatures, as suggested in~\cite{Kuroyanagi:2014nba}. We define the energy scales analogous to Eq.s~\eqref{eq:gwkdec}-\eqref{eq:gwkRHS},
\begin{align}
    \label{eq:gwkdecphi}
    k_\text{dec.}^\phi &= 1.7 \times 10^{14} \, \text{Mpc}^{-1} \left(\frac{g_{*S}(T_\phi)}{g_{*S}^0}\right)^{1/6} \left(\frac{T_\phi}{10^7 \, \text{GeV}}\right) \\
    \label{eq:gwkdecN}
    k_\text{dec.}^N &= 1.7 \times 10^{14} \, \text{Mpc}^{-1} \left(\frac{g_{*S}(T_{N_1})}{g_{*S}^0}\right)^{1/6} \left(\frac{T_{N_1}}{10^7 \, \text{GeV}}\right) \\
    k_\text{dec.S}^\phi &= k_\text{dec.}^\phi \Delta_\phi^{2/3} \\
    k_\text{dec.S}^N &= k_\text{dec.}^N \Delta_N^{2/3} \\
    \label{eq:2stepkRHS}
    k_\text{RH S}^\text{2-step} &= k_\text{RH} (\Delta_\phi\Delta_N)^{-1/3}
\end{align}
where $\Delta_\phi,\Delta_N$ are the dilution factors from $\phi$ and RHN decays respectively. Here in Eq.~\eqref{eq:2stepkRHS}, we utilized the fact that the total dilution factor after $\phi$ and RHN decay is the product of the individual dilution factors $\Delta_\phi$ and $\Delta_N$ corresponding to the two stages of entropy injection. A proper way to obtain the 2-step transfer function requires numerical fitting which we left as a future direction. In this case, we have two suppression frequencies corresponding to majoron and RHN decays, analogous to Eq.~\eqref{eq:fsup},
\begin{equation}
    \label{eq:fsup2step}
    f_{\rm sup}^\phi\simeq 2.7\times10^{-8}\;\text{Hz}\; \left(\frac{T_\phi}{{\rm GeV}}\right),\quad 
    f_{\rm sup}^N\simeq 2.7\times10^{-8}\;\text{Hz}\; \left(\frac{T_{N_1}}{{\rm GeV}}\right),\quad 
\end{equation}
And similar to Eq.~\eqref{eq:Rsup}, the suppression factors corresponding to the first ($\phi$ decay) and the second ($N$ decay) step of entropy injection are given by,
\begin{equation}
    \label{eq:Rsup2step}
    R_\text{sup}^{\rm 1st} \simeq \frac{1}{ (\Delta_\phi\Delta_N)^{4/3}},\quad R_\text{sup}^{\rm 2nd} \simeq \frac{1}{ \Delta_N^{4/3}}
\end{equation}


\subsection{Signal-to-noise ratio (SNR)} 
Interferometers are instruments that measure displacements in terms of dimensionless strain-noise denoted as $h_{\text{GW}}(f)$. This strain-noise is related to the amplitude of gravitational waves (GWs) and can be converted into an energy density, expressed by the formula,

\begin{equation}
    \Omega_{\text{exp}}(f)h^2 = \frac{2\pi^2 f^2}{3\mathcal{H}_0^2} h_{\text{GW}}(f)^2h^2
\end{equation}

Here, $\mathcal{H}_0$ represents the present-day Hubble rate ($\mathcal{H}_0 = h \times 100 \frac{\text{km/s}}{\text{Mpc}}$). To assess the likelihood of detecting the primordial GW background, we calculate the signal-to-noise ratio (SNR) using the experimental sensitivity $\Omega_{\text{exp}}(f)h^2$, either given or projected. The SNR is determined by the following formula,

\begin{equation}\label{eq:SNR}
    \text{SNR} \equiv \sqrt{\tau \int_{f_{\text{min}}}^{f_{\text{max}}} df \left(\frac{\Omega_{\text{GW}}(f)h^2}{\Omega_{\text{exp}}(f)h^2}\right)^2}
\end{equation}

In this analysis, we use $h = 0.7$ and an observation time of $\tau = 4$ years. A detection threshold of SNR $\geq 10$ is applied. It is important to note that this formula for SNR calculation is derived under the weak signal approximation, where the GW signal is much smaller than the instrumental noise~\cite{Allen:1997ad}. While it may overestimate the true SNR for strong signals, we adopt it here for both weak and strong GW signals for simplicity. 

In the spectrum plots, we display the sensitivity curves for different ongoing and future GW experiments. They can be grouped as: 
\begin{itemize}
    \item \textbf{Ground-based interferometers:} These detectors, such as \textsc{LIGO}/\textsc{VIRGO} \cite{LIGOScientific:2016aoc,LIGOScientific:2016sjg,LIGOScientific:2017bnn,LIGOScientific:2017vox,LIGOScientific:2017ycc,LIGOScientific:2017vwq}, a\textsc{LIGO}/a\textsc{VIRGO} \cite{LIGOScientific:2014pky,VIRGO:2014yos,LIGOScientific:2019lzm}, \textsc{AION} \cite{Badurina:2021rgt,Graham:2016plp,Graham:2017pmn,Badurina:2019hst}, \textsc{Einstein Telescope (ET)} \cite{Punturo:2010zz,Hild:2010id}, and \textsc{Cosmic Explorer (CE)} \cite{LIGOScientific:2016wof,Reitze:2019iox}, use interferometric techniques on the Earth's surface to detect gravitational waves.
    
    \item \textbf{Space-based interferometers:} Space-based detectors like \textsc{LISA} \cite{Baker:2019nia}, \textsc{BBO} \cite{Crowder:2005nr,Corbin:2005ny,Cutler:2009qv}, \textsc{DECIGO}, \textsc{U-DECIGO} \cite{Seto:2001qf,Yagi:2011wg}, \textsc{AEDGE} \cite{AEDGE:2019nxb,Badurina:2021rgt}, and \textsc{$\mu$-ARES} \cite{Sesana:2019vho} are designed to detect gravitational waves from space, offering different advantages over ground-based counterparts.
    
    \item \textbf{Recasts of star surveys:} Surveys such as \textsc{GAIA}/\textsc{THEIA} \cite{Garcia-Bellido:2021zgu} utilize astrometric data from stars to indirectly infer the presence of gravitational waves.
    
    \item \textbf{Pulsar timing arrays (PTA):} PTA experiments like \textsc{SKA} \cite{Carilli:2004nx,Janssen:2014dka,Weltman:2018zrl}, \textsc{EPTA} \cite{EPTA:2015qep,EPTA:2015gke}, and \textsc{NANOGRAV} \cite{NANOGRAV:2018hou,Aggarwal:2018mgp,NANOGrav:2020bcs} use precise timing measurements of pulsars to detect gravitational wave signatures.
    
    \item \textbf{CMB polarization:} Experiments like Planck 2018 \cite{Planck:2018jri,Planck:2018vyg}, BICEP 2/Keck \cite{BICEP:2021xfz} as computed by \cite{Clarke:2020bil}, and \textsc{LiteBIRD} \cite{Hazumi:2019lys} focus on detecting gravitational wave imprints in the polarization of the cosmic microwave background radiation.
    
    \item \textbf{CMB spectral distortions:} Projects like \textsc{PIXIE}, \textsc{Super-PIXIE} \cite{Kogut:2019vqh}, and \textsc{VOYAGER2050} \cite{Chluba:2019kpb} aim to detect gravitational wave signatures through spectral distortions in the cosmic microwave background.
\end{itemize}
The blue violin lines in the spectrum plots are the most recent data from NANOGrav.

\subsection{Bounds from BBN and CMB on dark radiation}
The gravitational wave energy density should be smaller than the bound on dark radiation parameterized as the number of additional neutrinos $\Delta N_{\rm eff}$~\cite{Luo:2020fdt,Maggiore:1999vm},
\begin{equation}
    \int_{f_{\rm min}}^{\infty}\frac{df}{f}\Omega_{\rm GW}(f)h^2\leq5.6\times10^{-6}\Delta N_{\rm eff}
\end{equation}
We neglect the frequency dependence of this bound and set $\Omega_{\rm GW}\leq 5.6\times10^{-6}\Delta N_{\rm eff}$ for the GW spectra that we calculate. BBN puts a bound on $\Delta N_{\rm eff}^{\rm BBN}\simeq 0.4$~\cite{Cyburt:2015mya}. Planck plus BAO observations set the bound $\Delta N_{\rm eff}^{\rm Plack+BAO}\simeq0.28$~\cite{Planck:2018vyg}. Projected bounds from future experiments are $\Delta N_{\rm eff}^{\rm Proj.}=0.014$ for CMB-HD~\cite{CMB-HD:2022bsz}, $\Delta N_{\rm eff}^{\rm Proj.}=0.05$ for CMB-Bharat~\cite{CMB-bharat}, $\Delta N_{\rm eff}^{\rm Proj.}=0.06$ for CMB Stage IV~\cite{doi:10.1146/annurev-nucl-102014-021908} and NASA's  PICO mission~\cite{Alvarez:2019rhd}, $\Delta N_{\rm eff}^{\rm Proj.}\lesssim0.12$ for CORE~\cite{CORE:2017oje}, the South Pole Telescope~\cite{SPT-3G:2014dbx} and Simons observatory~\cite{SimonsObservatory:2018koc}.

\begin{figure}[!ht]
    \centering
    \includegraphics[width=0.49\linewidth]{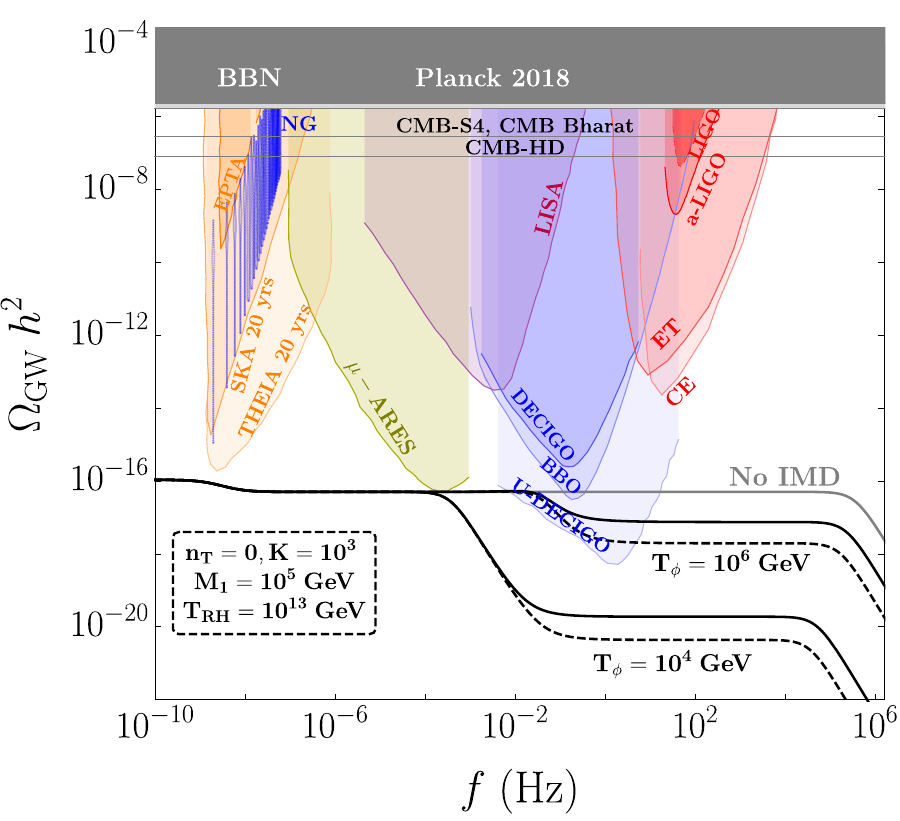} 
    \includegraphics[width=0.49\linewidth]{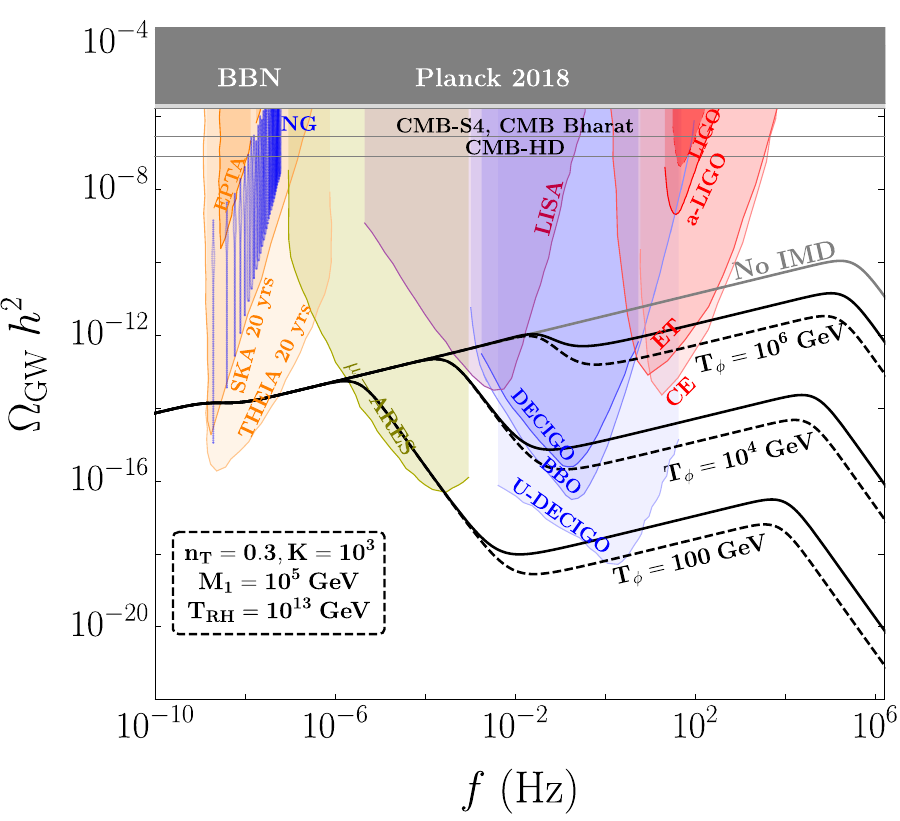} 
        \caption{\small \it Suppressed gravitational spectra of a benchmark Case (a) scenario when RHNs decay instantaneously or Case (b) where RHNs decay relativistically. Here we have taken $r=0.035$, $T_{\rm RH}=10^{13}$ GeV, $M_1=10^5$ GeV, $K=10^{3}$, and varied $T_\phi$ and $M_\phi$. Value of spectral index is $n_T = 0$ (\textbf{Left}) and $n_T = 0.3$ (\textbf{Right}). The gray line shows the standard GW spectrum without any intermediate matter domination. Black solid and dashed lines correspond to $M_\phi = 10^9$ and $3\times10^9$ GeV respectively. The extreme dampening at frequencies above $\sim100$ Hz is due to inflaton decay. \label{fig:nTH_CaseA}}
    \end{figure}

\begin{figure}[!ht]
\centering
\includegraphics[width=0.49\linewidth]{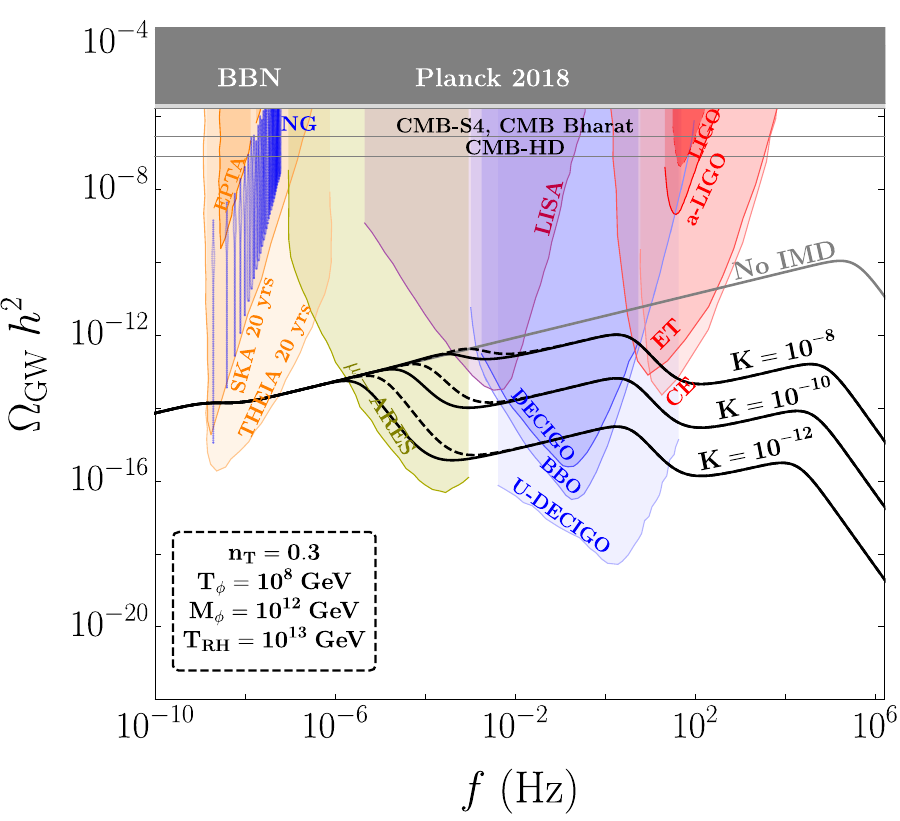} 
\includegraphics[width=0.49\linewidth]{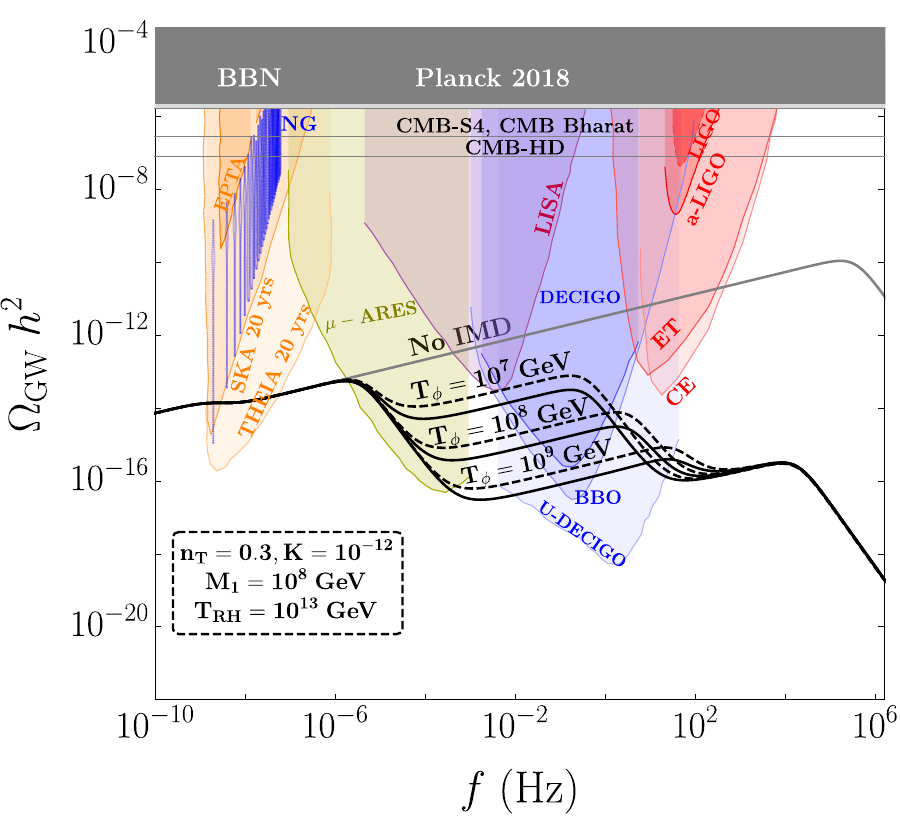} 
    \caption{\small \it Same as Fig. \ref{fig:nTH_CaseA} but for a benchmark Case (c), ``RHN matter domination" scenario. The \textbf{Left} plot shows the dependence of the spectra on $M_1$ and $K$ while the \textbf{Right} plot shows the dependence on $M_\phi$ and $T_\phi$. In the \textbf{Left} plot, the solid and dashed black lines represent $M_1=10^8$ and $3\times10^8$ GeV respectively. In the \textbf{Right} plot, the solid and dashed black lines represent $M_\phi=10^{12}$ and $2\times10^{12}$ GeV respectively.
    \label{fig:nTH_caseC}}
\end{figure}

\medskip 

\section{Results}\label{sec:results}
\subsection{Example spectra for different cases}\label{sec:resultsSpec}

Fig.~\ref{fig:nTH_CaseA} illustrates the damped gravitational wave (GW) spectra (black lines) for Case (a) and Case (b) discussed in Sec. \ref{sec:classification}, assuming  $r=0.035, M_1=10^5$ GeV, $K=10^3$ and spectral index $n_T=0$ (\textbf{Left}) and $n_T=0.3$ (\textbf{Right}). Here we demonstrate the influence of $T_\phi$ and $M_\phi$ on the GW spectra. We set the minimum $T_\phi$ to be approximately 100 GeV to accommodate sphaleron effects within the electroweak scale. The gray solid line marked ``No IMD'' is the standard spectrum with no intermediate matter-dominated phase while the black solid and dashed lines are the spectral shapes in the presence of the matter-dominated epoch. Different colored regions correspond to sensitivity curves of current and future GW experiments. The blue violin lines are taken from the recent NANOGrav 15yr dataset~\cite{NANOGrav:2023hvm,NANOGrav:2023gor}. Various horizontal lines on the top of the plot are coming from the current or projected $\Delta N_{\rm eff}$ bounds from BBN, Planck 2018, and CMB surveys~\cite{Berbig:2023yyy}. Higher frequencies in the figure correspond to earlier times, showcasing suppression beyond the critical frequency $f_{\rm sup}$ due to entropy injection given by Eq.~\eqref{eq:fsup}. Suppression at frequencies higher than $f_{\rm sup}\sim 10^{-5}-0.1$ Hz are attributed to majoron and RHN decay, while suppression at frequencies higher than $\sim 10$ Hz is a consequence of inflaton decay terminating at the reheating temperature $T_{\rm RH}=10^{13}$ GeV. Lower values of $T_\phi$ shift the suppression frequency to the left and increase the dilution factor as expected from Eq.~\eqref{eq:CaseA_Delta}. The solid and dashed spectral lines represent $M_\phi=10^9$ and $3\times10^9$ GeV respectively. A higher $M_\phi$ means more dilution. Importantly, GW signatures remain invariant across weak and strong washout regimes, as they solely depend on the majoron dynamics in this case.

\begin{figure}[!ht]
\centering
    \begin{subfigure}{0.5\linewidth}
\includegraphics[width=\linewidth]{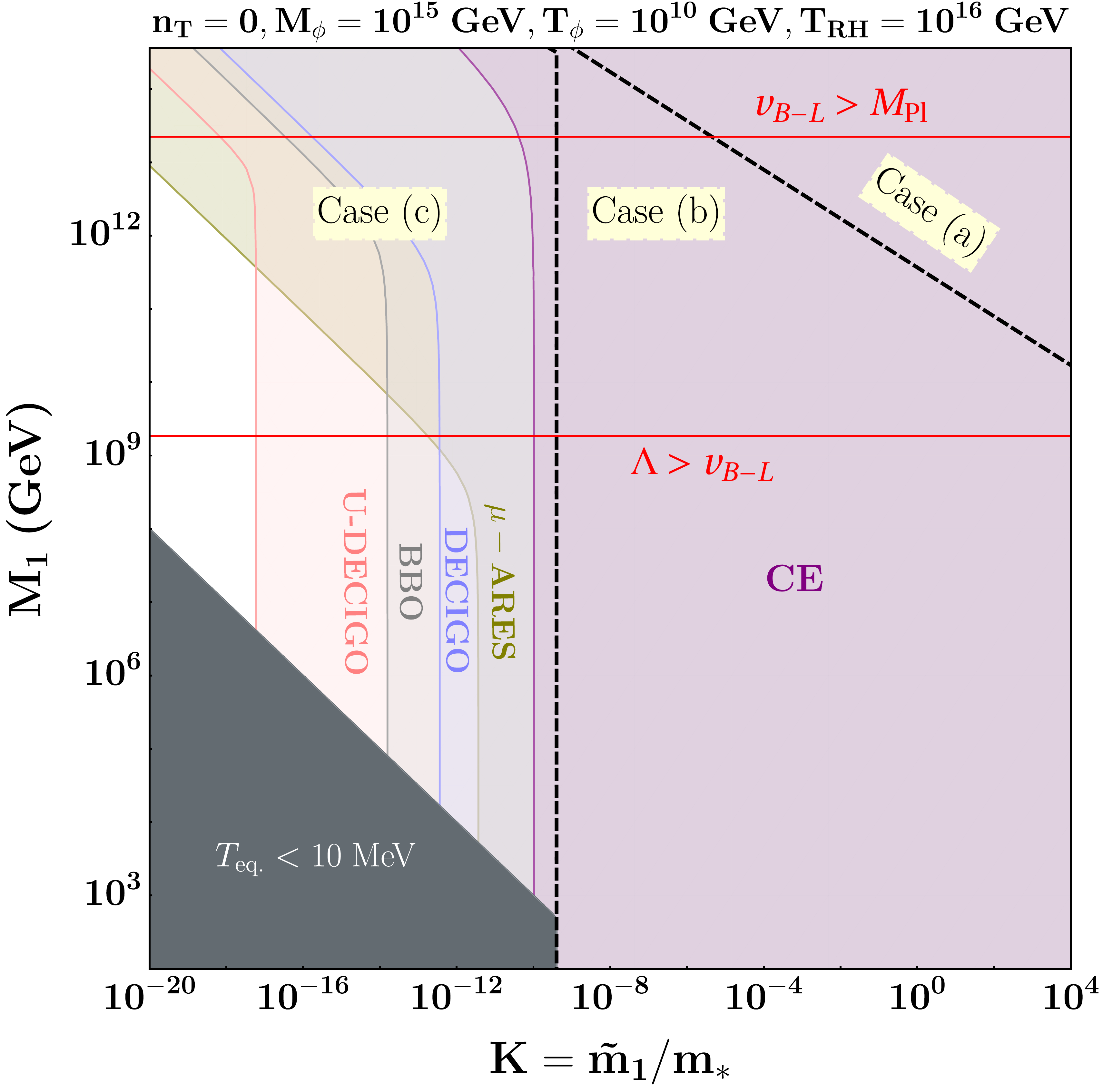} 
    \label{fig:SNR0M15T10}
    \end{subfigure}\hfill
    \begin{subfigure}{0.5\linewidth}
\includegraphics[width=\linewidth]{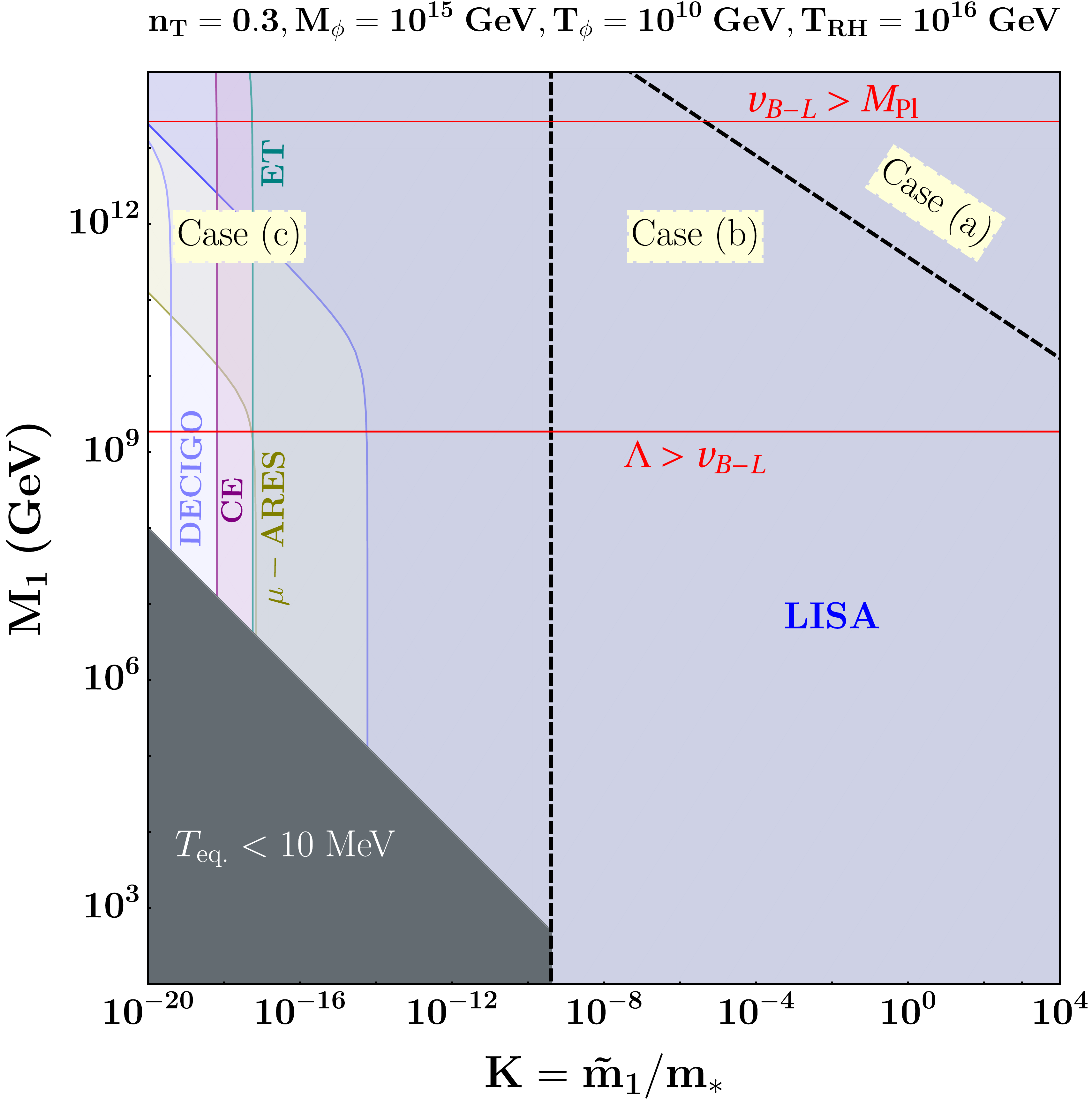}
    \label{fig:SNR05M15T10}
    \end{subfigure}
    
\caption{\small \it Projected sensitivities of different experiments in the RHN parameter space for $n_T=0$ (\textbf{Left}) and $n_T=0.3$ (\textbf{Right}). Here $\tilde{m}_1$ is the effective neutrino mass defined by Eq.~\eqref{eq:effneutrinomass} and $m_*=1.1\times10^{-3}$ eV is the equilibrium neutrino mass defined in Eq.~\eqref{eq:eqneutrinomass}. For both plots we took $M_\phi$= $10^{15}$ GeV , $T_\phi$ = $10^{10}$ GeV, $r=0.035$. The colored regions associated with each experiment imply sensitivity with SNR $>10$. As expected, a higher $n_T$ increases overall SNR. Non-thermal leptogenesis is possible in this model only in the region between the two solid red lines. See the main text for a detailed description and analysis.}
    \label{fig:SNRM15T10n}
\end{figure}

Fig.~\ref{fig:nTH_caseC} presents the GW signature of long-lived RHNs in Case (c), depicting the two stages of entropy injection. We show the dependence of the spectra on parameters $K$ and $M_1$ (\textbf{Left}) and on $M_\phi$ and $T_\phi$ (\textbf{Right}). The necessary condition for this case is $\sqrt{K}<2T_\phi/M_\phi$. We choose the combination $T_{N_1} = M_1 \sqrt{K}$ in a way such that $T_{N_1}\geq100\;\text{GeV}$. For these spectra, we take  $r=0.035$, $n_T=0.3$, and use the 2-step transfer function utilizing Eq.~\eqref{eq:2stepF}. Notably, the majoron's effect manifests as an additional dip in the spectrum, a knee-like feature, positioned between the dips induced by inflaton decay and RHN decay. Thus we have two knee-like features, one from $\phi$ decay and another from RHN decay. In the \textbf{Left} plot, the solid and dashed spectral lines correspond to $M_1=10^8$ and $3\times10^8$ GeV respectively. Here we take $M_\phi=10^{12}$ GeV, $T_\phi=10^8$ GeV and $T_{\rm RH}=10^{13}$ GeV. We see that a lower $K$ and $M_1$ lead to higher suppression, as expected from Eq.~\eqref{eq:CaseC_Delta}. In the \textbf{Right} plot, we fix $M_1=10^8$ GeV, $K=10^{-12}$, $T_{\rm RH}=10^{13}$ GeV. Here solid and dashed spectral lines correspond to $M_\phi=10^{12}$ and $2\times10^{12}$ GeV respectively. Here, we see that the overall suppression increases with increasing $T_\phi$ and decreasing $M_\phi$, also understood from Eq.~\eqref{eq:CaseC_Delta}. This trend is opposite to the trend in Cases (a), (b) as shown in Fig.~\ref{fig:nTH_CaseA}.

\begin{figure}[!ht]
    \centering
        \begin{subfigure}{0.5\linewidth}
    \includegraphics[width=\linewidth]{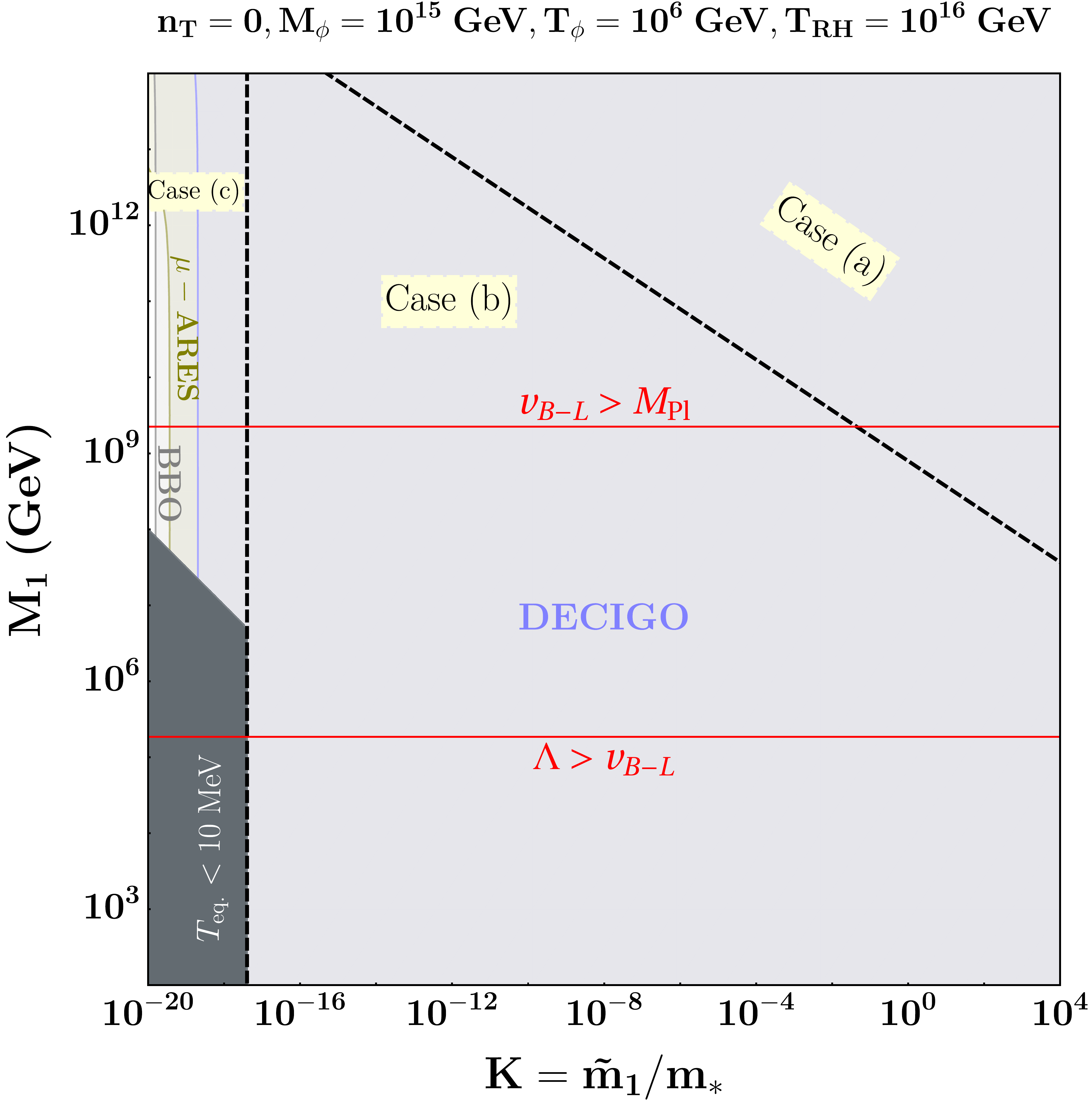} 
        \label{fig:SNR0M15T6}
        \end{subfigure}\hfill
        \begin{subfigure}{0.5\linewidth}
    \includegraphics[width=\linewidth]{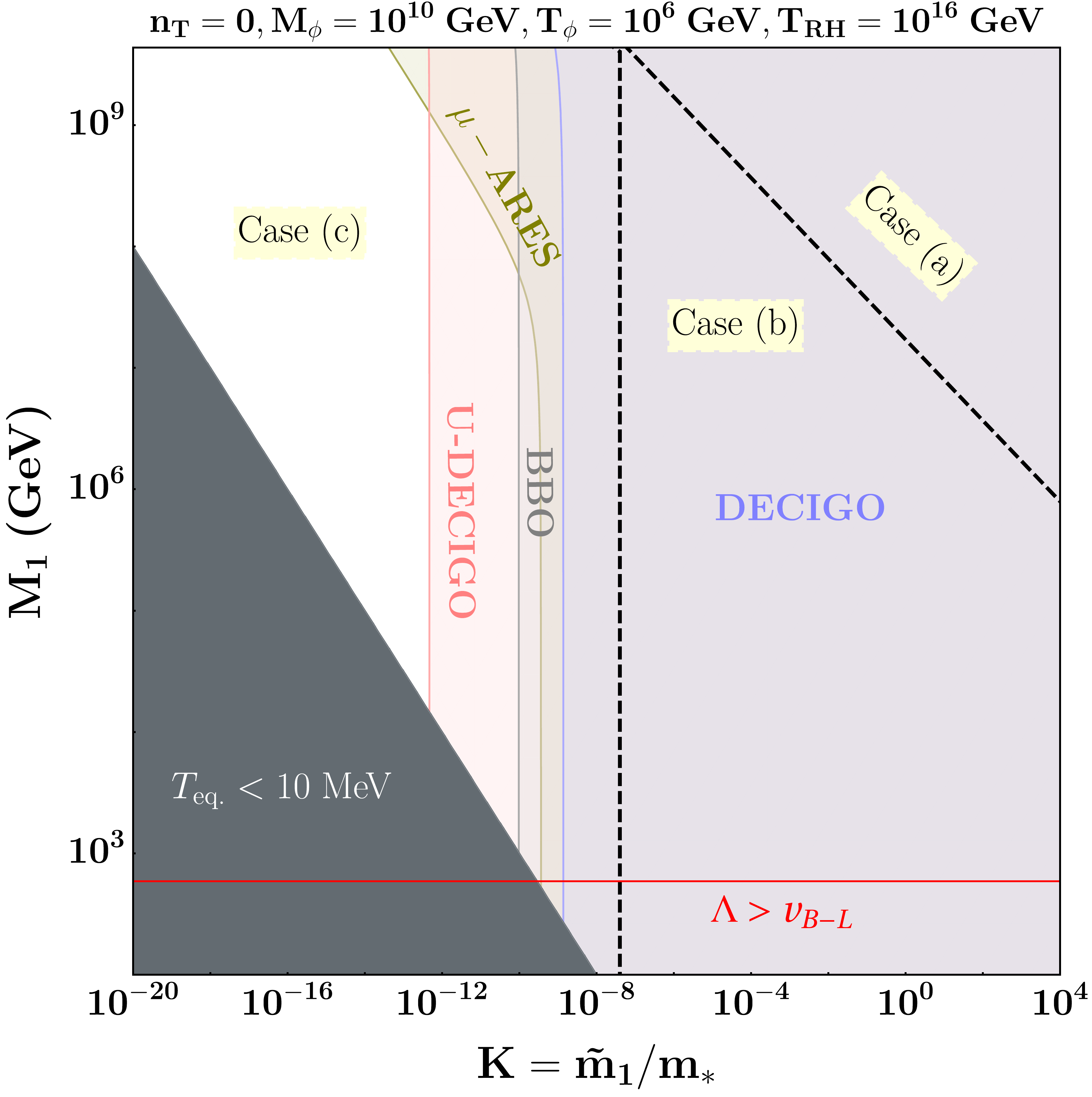}
        \label{fig:SNR0M10T6}
        \end{subfigure}
        
    \caption{\small \it Same as Fig. \ref{fig:SNRM15T10n} but for $n_T=0$, $T_\phi$ = $10^6$ GeV and $M_\phi$= $10^{15}$ GeV (\textbf{Left}) and $10^{10}$ GeV (\textbf{Right}). For a fixed $T_\phi$, the left plot is expected to have smaller overall SNR values due to higher $M_\phi$. Here $m_*=1.1\times10^{-3}$ eV.}
        \label{fig:SNRn0T6M}
    \end{figure}

In Case (c), the suppression factor associated with the the middle ``knee-like'' feature is given by, $R^{\rm 1st}_{\rm sup}=(\Delta_\phi\Delta_N)^{-4/3}$, defined in Eq.~\eqref{eq:Rsup2step}. Minimum possible value of this factor corresponds to the maximum suppression. In this case, $R^{\rm 1st}_{\rm sup}\sim 693.5\,K^{2/3}$, obtained by substituting Eq.~\eqref{eq:CaseA_Delta} and~\eqref{eq:CaseC_Delta} in Eq.~\eqref{eq:Rsup2step}. Minimum value of the washout parameter $K$ is obtained by considering RHN decay temperature $T_{N_1}=M_1\sqrt{K}\sim100$ GeV such that sphalerons can successfully convert lepton asymmetry to baryon asymmetry. For maximum value of $M_1=10^{14}$ GeV, we obtain the minimum value of $R^{\rm 1st}_{\rm sup}\simeq 7\times10^{-14}$. If we do not consider leptogenesis, then $T_{N_1}$ can go as low as BBN temperature which is around 10 MeV. In that case the minimum suppression factor is $R^{\rm 1st}_{\rm sup}\simeq3\times10^{-19}$. If in the future, two-step entropy injection with a smaller suppression factor is detected, it would rule out non-thermal leptogenesis scenario within our model.

\subsection{Estimates for Signal to Noise Ratio}
\begin{figure}[!ht]
    \centering
    \includegraphics[width=0.5\linewidth]{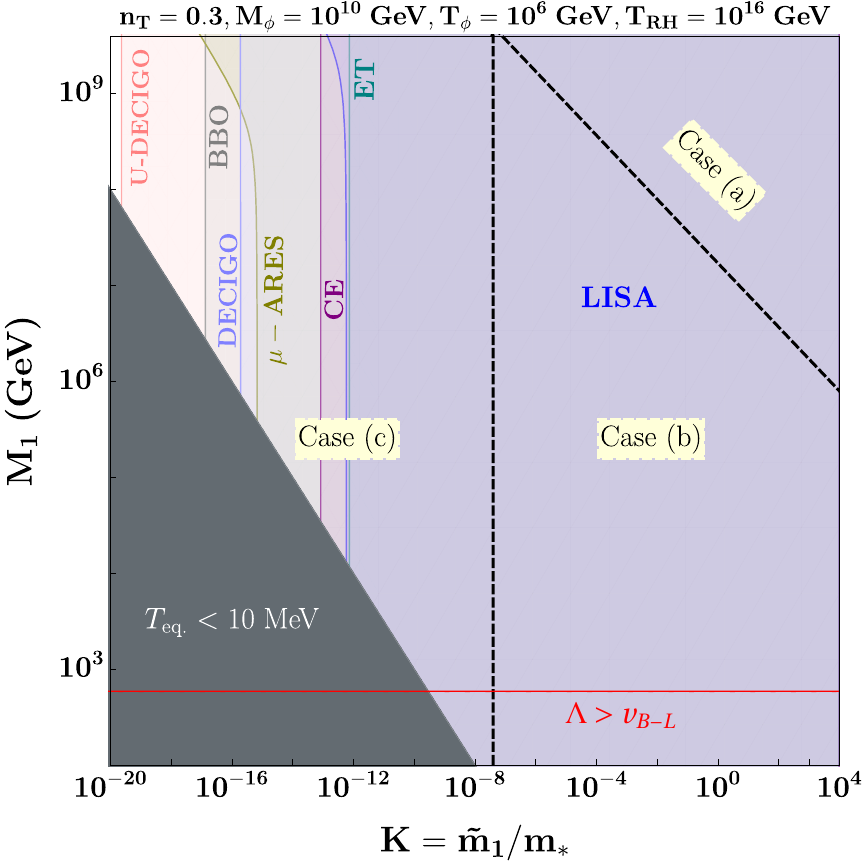} 
        \caption{\small \it Same as Fig. \ref{fig:SNRn0T6M} (\textbf{Right}) but for $n_T=0.3$. Here $m_*=1.1\times10^{-3}$ eV.\label{fig:nTH_SNR03M10T6}}
    \end{figure}
\begin{figure}[!ht]
\centering
\includegraphics[width=0.5\linewidth]{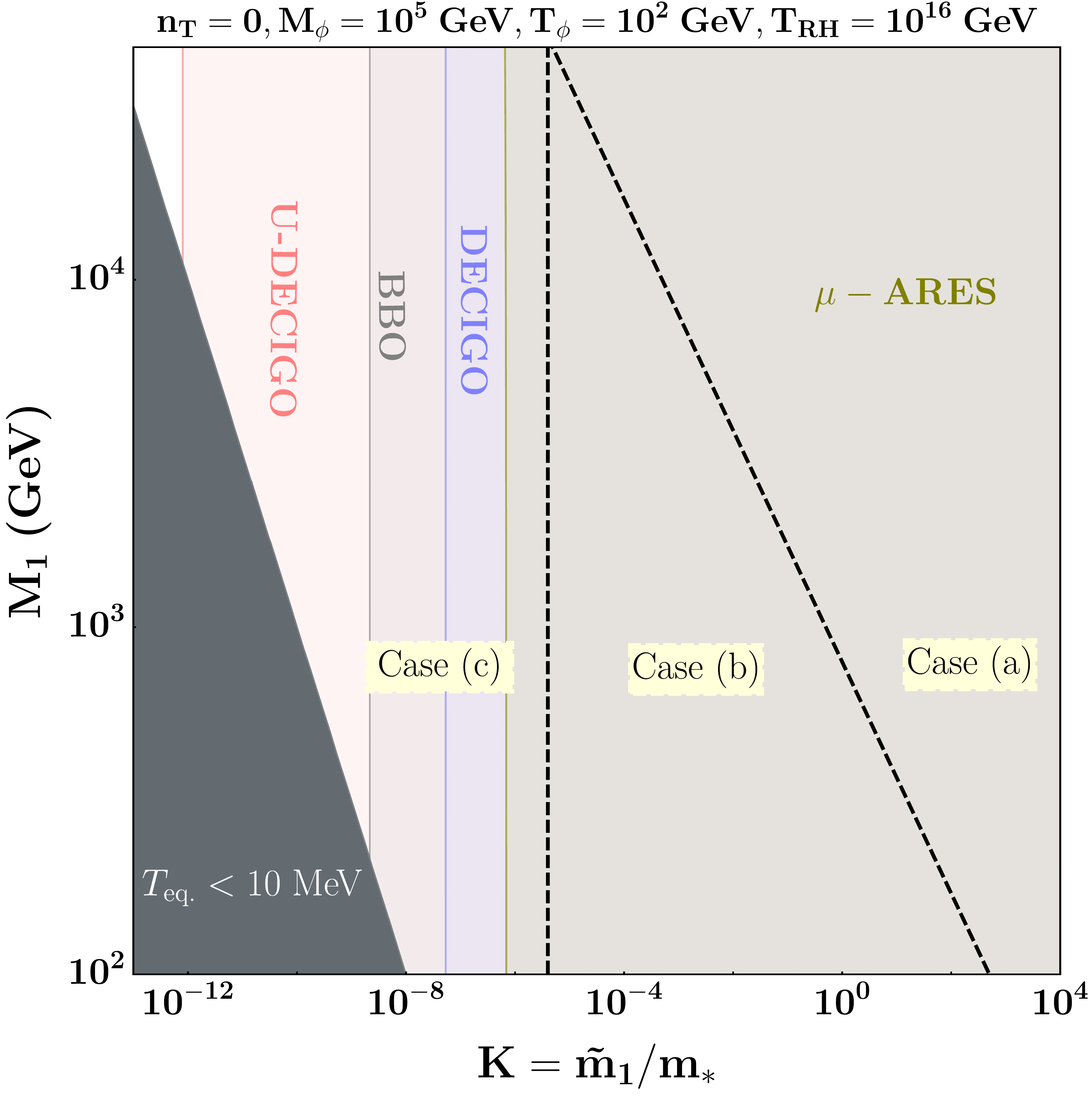} 
    \caption{\small \it Same as Fig. \ref{fig:SNRM15T10n} but for $T_\phi=100$ GeV and $n_T=0$, $M_\phi=10^5$ GeV. Here $m_*=1.1\times10^{-3}$ eV.\label{fig:SNRM5}}
\end{figure}

The next set of Figures~\ref{fig:SNRM15T10n} -~\ref{fig:SNRM5} demonstrates the projected future sensitivities of U-DECIGO, DECIGO, BBO, $\mu$-ARES, LISA, CE and ET experiments in terms of signal to noise ratio (SNR) given in Eq.~\eqref{eq:SNR}, in the parameter space of the RHN for different values of $n_T,M_\phi,T_\phi,T_{\rm RH}$. In these figures, the black dashed lines separate the Case (a), Case (b) and Case (c) regions. In each of the SNR plots, the gray-shaded region is excluded due to the requirement that the latest radiation-dominated era should start before BBN takes place at around temperatures of a few MeV. In our model, the latest radiation domination starts at the equilibrium temperature $T_{\rm eq.}=T_{\rm dec.}$ defined in Eq.~\eqref{eq:Teq}. In all the SNR plots, the colored regions associated with different experiments can be probed by that experiment with SNR $\ge 10$. In all these SNR plots we choose $T_{\rm RH}=10^{16}$ GeV to eliminate its effect on the SNR. The two solid red lines represent theoretical constraints on our model. In the region above the upper red line, $v_{B-L}$ becomes larger than Planck scale while below the lower red line, the requirement that the confinement scale should be less than the $B-L$ breaking scale, $\Lambda < v_{B-L}$, discussed in Sec.~\ref{sec:lepscalar}, is not satisfied. Hence, successful leptogenesis is possible only in the region between these two lines in our model.
In vanilla leptogenesis scenario, vacuum stability of SM Higgs gives an upper bound on the lightest RHN mass $M_1\lesssim10^{14}$ GeV~\cite{Casas:1999cd, Elias-Miro:2011sqh} and we assume it would be approximately the same in our model~\cite{Fu:2023nrn}. Hence we take the upper limit on $M_1$ to be around $10^{14}$ GeV. On the other hand, the perturbativity limit of the model requires the Yukawa couplings $\lambda_{ij}<4\pi$, which along with the necessity to produce at least one mass eigenstate $m_\nu=0.05$ eV gives another bound, $M_2\lesssim 10^{15}$ GeV. Hence, in our quasi-degenerate RHN model, we find the perturbativity bound $M_1\lesssim10^{15}$ GeV. The perturbativity limit of the light-heavy neutrino mixing $y_{N_i}< 4\pi$ is always satisfied in our model.

In Fig. \ref{fig:SNRM15T10n}, we pick a high majoron mass $M_\phi=10^{15}$ GeV, and decay temperature $T_\phi=10^{10}$ GeV and show the projected sensitivities of different experiments for $n_T=0$ (\textbf{Left}) and $n_T=0.3$ (\textbf{Right}). We see that increasing the spectral index $n_T$ also increases the GW-sensitive region by increasing the SNR as expected. For $n_T=0$, LISA and ET will not be able to probe leptogenesis but for $n_T=0.3$ both the experiments will be sensitive to a large portion of the parameter space. In the Case (a) and Case (b) regions, the SNR will not depend on $M_1$ and $K$ but only on $M_\phi$ and $T_\phi$. In the Case (c) region, RHNs decay at $T_{N_1}=M_1\sqrt{K}$. Hence decreasing $M_1$ or $K$ results in more suppression and lower SNR. 


In Fig.~\ref{fig:SNRn0T6M}, we show the sensitivities for $n_T=0$, $T_\phi$ = $10^6$ GeV and $M_\phi$= $10^{15}$ GeV (\textbf{Left}) and $10^{10}$ GeV (\textbf{Right}). We see that for $M_\phi = 10^{10}$ GeV, the GW sensitive region in the parameter space is smaller compared to that for $M_\phi = 10^{15}$ GeV. Also, comparing Fig.~\ref{fig:SNRn0T6M} (\textbf{Left}) with Fig.~\ref{fig:SNRM15T10n}, we see that the Case (a) and Case (b) regions cover more space for lower $T_\phi$ values. We notice that CE is no longer sensitive for $n_T=0, T_\phi=10^6$ GeV in Fig.~\ref{fig:SNRn0T6M} (\textbf{Left}) while it is sensitive for $T_\phi=10^{10}$ GeV in Fig.~\ref{fig:SNRM15T10n} (\textbf{Left}). 

In Fig. \ref{fig:nTH_SNR03M10T6}, we show the SNR for $n_T=0.3$, $T_\phi=10^6$ GeV and $M_\phi=10^{10}$ GeV. Comparing with Fig.~\ref{fig:SNRn0T6M} (\textbf{Right}) plot, again the effect of $n_T$ is seen, and LISA, CE, and ET will be able to probe the parameter space for $n_T=0.3$.

Fig.~\ref{fig:SNRM5} shows the shape of the sensitive regions for $M_\phi=10^5$ GeV and $T_\phi=100$ GeV assuming $n_T=0$. We see that UDECIGO, DECIGO, BBO, $\mu-$ARES can probe leptogenesis with $T_\phi=100$ GeV, given small mass $M_\phi$ of the majoron. A small majoron mass injects less amount of entropy when they decay as a result of which the suppression is small, hence SNR is larger.

We calculate the SNR contours for all these GW experiments taking various values of $M_\phi$ and $T_\phi$ for $n_T=0$ and $0.3$ but do not show all of them in this paper. However, based on these results and Fig.~\ref{fig:SNRM15T10n}-\ref{fig:SNRM5}, we summarize which key GW experiments can probe the parameter space $M_1$ vs $K$ in Tables~\ref{tab:SNRexpt0}.

\begin{table}[htbp]
    \centering
    \tiny
    \begin{tabular*}{\linewidth}{@{\extracolsep{\fill}}|c|c|c|c|c|c|c|c|c|c|c|c|c|c|}
        \hline
        \hline
        $M_\phi$ (GeV) & $T_\phi$ (GeV) &
        \multicolumn{2}{c|}{U-DECIGO} &
        \multicolumn{2}{c|}{BBO} &
        \multicolumn{2}{c|}{$\mu$-ARES} &
        \multicolumn{2}{c|}{LISA} &
        \multicolumn{2}{c|}{ET} &
        \multicolumn{2}{c|}{CE} \\
        \cline{3-14}
        & & $n_T=0$ & $n_T=0.3$ & $n_T=0$ & $n_T=0.3$ & $n_T=0$ & $n_T=0.3$ & $n_T=0$ & $n_T=0.3$ & $n_T=0$ & $n_T=0.3$ & $n_T=0$ & $n_T=0.3$ \\
        \hline

        \multirow{3}{*}{$10^{15}$}
            & $10^{10}$ & \checkmark & \checkmark & \checkmark & \checkmark & \checkmark & \checkmark & -   & \checkmark & -   & \checkmark & \checkmark & \checkmark \\
            & $10^{6}$  & \checkmark & \checkmark & \checkmark & \checkmark & \checkmark & \checkmark & -   & \checkmark & -   & \checkmark & -   & \checkmark \\
            & $10^{2}$  & -   & -   & -   & -   & -   & \checkmark & -   & -   & -   & -   & -   & -   \\
        \hline

        \multirow{2}{*}{$10^{10}$}
            & $10^{6}$  & \checkmark & \checkmark & \checkmark & \checkmark & \checkmark & \checkmark & -   & \checkmark & -   & \checkmark & -   & \checkmark \\
            & $10^{2}$  & -   & \checkmark & -   & \checkmark & -   & \checkmark & -   & -   & -   & -   & -   & -   \\
        \hline

        $10^{5}$ & $10^{2}$  & \checkmark & \checkmark & \checkmark & \checkmark & \checkmark & \checkmark & -   & \checkmark & -   & \checkmark & -   & \checkmark \\
        \hline
        \hline
    \end{tabular*}
    \caption{\small \it Future sensitivity of different key GW experiments to probe the RHN parameter space ($M_1$ vs. $K$) with threshold SNR $>10$, for benchmark values of $M_\phi$ and $T_\phi$. Columns show sensitivities for $n_T = 0$ and $n_T = 0.3$. Here the check mark implies that the particular experiment will be sensitive to non-thermal leptogenesis for given value of $n_T$. Dashes imply that in the cosidered parameter space, SNR $< 10$ for the given experiment. We considered $T_{\rm RH} = 10^{16}$ GeV and maximal tensor-to-scalar ratio $r = 0.035$.}
    \label{tab:SNRexpt0}
\end{table}

It is worth mentioning that there are certain models of inflation that go beyond the standard power-law description of the IGW spectrum. For example, a broken power-law spectrum of IGW is suggested in~\cite{Jiang:2023gfe}. The power-law is broken at a scale $k_{\rm break}>k_*$ corresponding to the frequency $f_{\rm break} \ll 10^{-8}$ Hz, where $k_*$ is the pivot scale defined in Eq.~\eqref{eq:PTprime}. Below $f_{\rm break}$, the spectrum is red-tilted ($n_T^{\rm CMB} \lesssim 0$), while above $f_{\rm break}$ it can be blue-tilted ($n_T > 0$). Since our focus is on $f \gg 10^{-8}$ Hz, we can replace $k_*$ with $k_{\rm break}$ in Eq.\eqref{eq:PTprime} to obtain broken power law assuming $n_T^{\rm CMB}=0$. In this case, $\Omega_{\rm GW}(k)$ and the resulting SNR are suppressed by a factor $(k_*/k_{\rm break})^{n_T} \sim \mathcal{O}(0.1$–$1)$ relative to the pure power-law case. Similarly, inflationary models with a Gauss-Bonnet term can yield a blue-tilted spectrum with a running $n_T$~\cite{Koh:2018qcy}, also modifying the SNR. However, the effect of an intermediate matter-dominated era quantified by the ratio of SNR with and without the matter-domination, remains unaffected. We therefore do not consider such extensions further in this work.

\medskip

\section{Discussion \& Conclusion}\label{sec:conclusion}

Leptogenesis at low and intermediate energy scales provides an attractive scenario for baryogenesis, linking the observed matter-antimatter asymmetry to the neutrino sector and seesaw scale. In this analysis, we offer a probe of this scenario via observations of the stochastic GW background that originates during inflation (which is responsible for the creation of primordial density perturbation of the Universe) and propagates in the post-inflationary era to reach us at present. Particularly, we constructed a global $U(1)_{B-L}$ model of neutrino mass that facilitates non-thermal leptogenesis through the decay of the Goldstone boson called majoron, denoted by $\phi$. We showed that detecting a characteristic GW spectral shape will provide an indication of a RHN or $\phi$ matter-dominated era in the early Universe. We present a minimal model realization of non-thermal leptogenesis, for which we found that a large fraction of the viable parameter space will be probed in future GW experiments. For the lightest RHN mass $M_1$ and washout parameter $K$ involving light-heavy neutrino Yukawa couplings $\lambda$, we find that $M_1$ in the range $100$ GeV $\lesssim M_1\lesssim 10^{14}$ GeV and $K$ for extremely weak washout ($K\sim10^{-20}$) and for strong washout ($K\sim10^4$) regime consistent with successful baryogenesis, novel GW shapes will be probes by U-DECIGO, DECIGO, BBO, $\mu-$ARES, LISA, CE, ET etc. By estimating signal-to-noise ratio (SNR) for GW experiments, we investigated the effect of the majoron mass $M_\phi$ and decay temperature $T_\phi$ which depends on the $\phi-N$ Yukawa couplings $y_N$.\\

\noindent The key interpretations of our results are summarized below:
\begin{itemize}
    \item The overall SNR is larger for a larger spectral index $n_T$ (compare for example Fig.~\ref{fig:SNRM15T10n} \textbf{Left} and \textbf{Right} plots).

    \item Non-thermal leptogenesis can be probed in future GW experiments such as U-DECIGO, BBO etc. (see spectra in Fig.~\ref{fig:nTH_CaseA}-\ref{fig:nTH_caseC} and SNR in Fig.~\ref{fig:SNRM15T10n} - \ref{fig:SNRM5}).

    \item If $T_{N_1}\ll T_\phi$ in Case (c) scenario with 2-step entropy injection, lower values of $M_1$ and $K$ reduce SNR and therefore challenging to test for all experiments (Fig.~\ref{fig:SNRM15T10n}-\ref{fig:SNRM5}).
    
    \item For Case (a) and Case (b) scenarios discussed in Sec.~\ref{sec:classification}, a lower $T_\phi$ value decreases the SNR and frequency of suppression $f_{\rm sup}$ given in Eq.~\eqref{eq:fsup}, making it more difficult to observe (see spectra in Fig.~\ref{fig:nTH_CaseA}). However, suitable low values of $T_\phi$ increase the overall GW sensitive region in the parameter space of the lightest RHN (compare for example Fig.~\ref{fig:SNRM15T10n} and \ref{fig:SNRn0T6M}).
    
    \item A higher $M_\phi$ in general means larger entropy injection which decreases the overall SNR values for all experiments (compare solid and dashed spectra in Fig.~\ref{fig:nTH_CaseA}). However, for Case (c), higher $M_\phi$ increases the SNR (solid and dashed spectra in Fig.~\ref{fig:nTH_caseC} (\textbf{Right})). $M_\phi$ also sets the upper bound on $M_1$ in our model, i.e. $M_1\leq M_\phi/2$. We are interested in the parameter space where $\phi$ is long-lived to dominate the energy Budget of the Universe (see discussion following Eq.~\eqref{eq:Tdom} and Eq.~\eqref{eq:phi_dom_condition}).

    \item In Case (a) and Case (b) described in Sec.~\ref{sec:classification}, the leptogenesis scale is $\sim T_\phi$ which means GW experiments can probe leptogenesis even for strong washout $K>1$ where RHNs might eventually thermalize with the radiation bath (Fig.~\ref{fig:SNRM15T10n}-\ref{fig:SNRM5}). This is unlike the model in~\cite{Berbig:2023yyy} where the RHN domination criterion $K\lesssim4\times 10^{-4}$ must be satisfied for GW sensitivity.

    \item Table~\ref{tab:SNRexpt0} summarizes which key experiments can probe non-thermal leptogenesis for $n_T=0$ and $0.3$ respectively, given benchmark values of the majoron mass $M_\phi$ and decay temperature $T_\phi$.

\end{itemize}

If the characteristic features of the GW spectral shapes proposed in this study are observed, one may look to target additional observations to distinguish between an RHN or majoron $\phi$ dominated pre-BBN era and other forms of early matter domination. Particularly in low-scale leptogenesis RHN masses of GeV-TeV could be searched in typical Heavy Neutral Lepton Searches (HNL) search experiments (see~\cite{Beacham:2019nyx,Abdullahi:2022jlv,Middleton:2022dio} for current experimental limits) at the particle physics laboratories. In this manner, we can complement GW searches with laboratory searches in the same BSM parameter space. However, such a study is beyond the reach of the present paper and we plan to explore this in future. RH neutrinos, if they exist should also show up in experiments such as neutrino-less double beta decay~\cite{Dolinski:2019nrj,Gomez-Cadenas2023} or lepton number violating processes~\cite{Li:2021fvw,Cai:2017mow} and therefore provide us with a myriad of pathways to independently verify the existence of an early RHN-domination or majoron domination era. This leads to a unique and exciting opportunity to form synergies between GW searches and laboratory and CMB searches.

\medskip

\acknowledgments
ZAB is supported by a DST-INSPIRE fellowship. LM is supported by a UGC fellowship. The authors thank Prof. Kai Schmitz for his valuable comments.

\appendix
\section{Structure of complex orthogonal matrix $R$}\label{app:C}
\begin{figure}[!ht]
\centering
\includegraphics[width=0.6\linewidth]{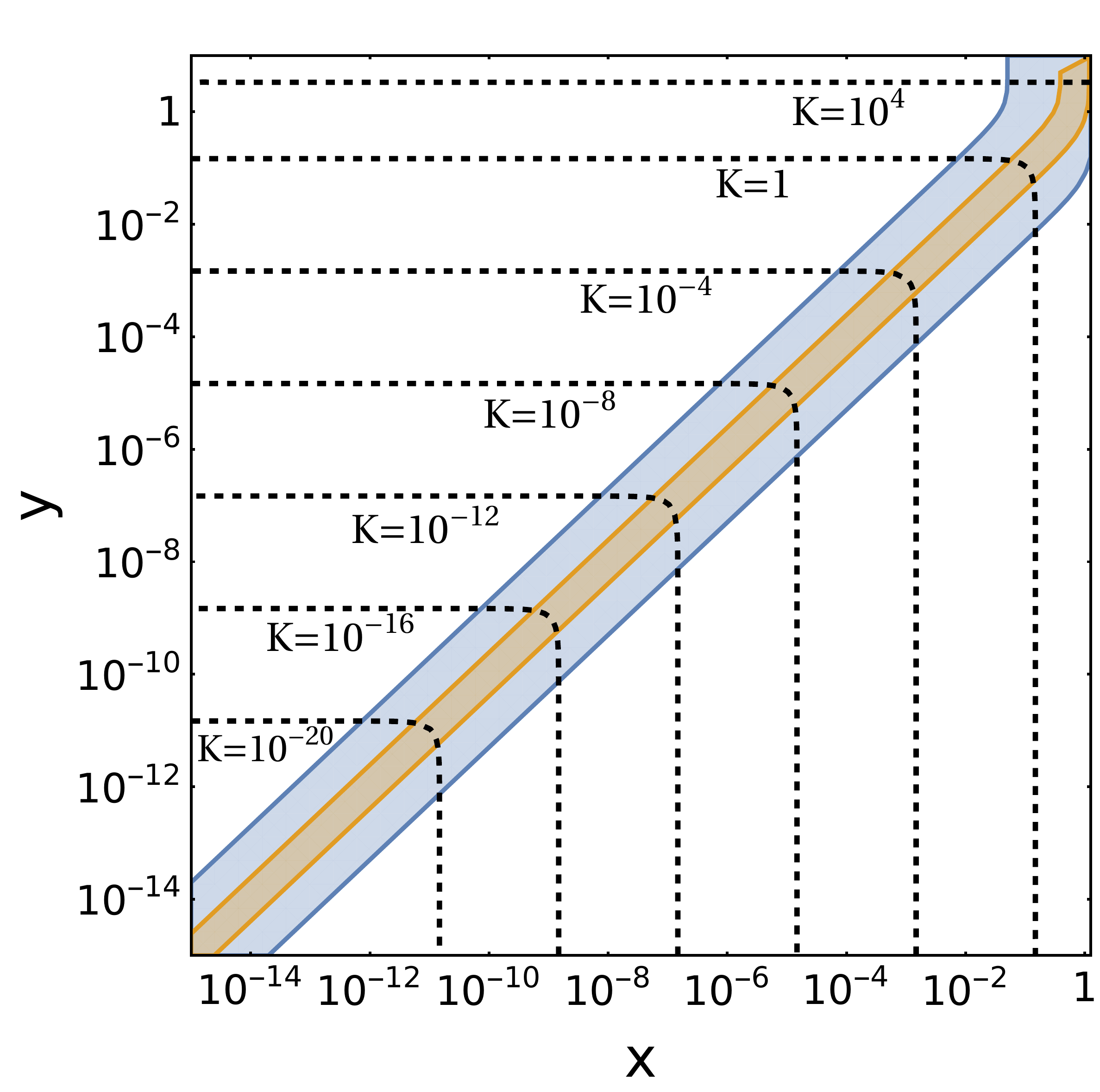}  
\caption{\small \it CP parameter $|\epsilon_1|$ defined in Eq.~\eqref{eq:epsilon2} and contours of decay parameter $K$ defined in Eq.~\eqref{eq:Kdefinition} in the paramter space spanned by $x$ and $y$. Blue and orange region signifies $|\epsilon_1|>10^{-7}$ and $7\times10^{-7}$ respectively. It is seen that for $x=y\ll1$, large $|\epsilon_1|$ can be obtained even for a tiny value of $K$.}
    \label{fig:CPplot}
\end{figure}

Here we explicitly show that CP asymmetry defined in Eq.~\eqref{eq:epsilon2} can be large even for a tiny washout parameter $K$ defined in Eq.~\eqref{eq:Kdefinition}. For example, take a simple form of $R$ defined in Eq.~\eqref{eq:casas},
\begin{equation}\label{eq:Rmatrix}
    R=\begin{pmatrix}
        \cos\omega & 0 & -\sin\omega \\
        0 & 1 & 0 \\
        \sin\omega & 0 & \cos\omega \\
    \end{pmatrix},
\end{equation}
where $\omega=x+iy$ is a complex angle. Assuming normal ordering and $m_1=0$, the CP asymmetry can be written in terms of $x$ and $y$ as,
\begin{equation}
    |\epsilon_1|=\frac{3}{32\pi}\frac{M_1 m_3}{v^2}\frac{\sin 2x\sinh 2y}{\sin^2x\cosh^2y+\cos^2x\sinh^2y}=\frac{3}{16\pi}\frac{M_1 m_3}{v^2}\frac{\sin 2x\sinh 2y}{\cosh 2y-\cos 2x}.
\end{equation}
When $x=y\ll1$, it can be verified that $|\epsilon_1|=\epsilon_1^{\rm max}$, which is the maximal CP asymmetry defined in Eq.~\eqref{eq:ep1max}. Taking $x=y=10^{-11}$ gives $|\epsilon_1|\sim10^{-6}$ (assuming $M_1=10^{10}$ GeV) while $K\sim10^{-20}$.
In Fig.~\ref{fig:CPplot} we show the regions where $|\epsilon_1|>10^{-7}$ (blue) and $|\epsilon_1|>7\times10^{-7}$ (orange) in the parameter space spanned by $x$ and $y$. Along with it, we also show the contours of constant $K$ values in the same parameter space. For this plot we assume normal ordering and $m_1=0$, $M_1=10^{10}$ GeV, $M_2=10^{11}$ GeV, $M_3=10^{12}$ GeV. We see that for such high RHN mass scales, it is possible to generate $|\epsilon_1|=\epsilon_1^{\rm max}$ even if $\tilde{m}_1\rightarrow0$ hence $K\rightarrow0$. However $K_2,K_3$ remains large.

\section{Change of Variable}
\label{app:A}
It is convenient to use comoving variables while solving the Boltzmann equations. We adopt the same prescription as in~\cite{Chung:1998rq,Hahn-Woernle:2008tsk},
\begin{equation}
    E_{\phi,N_i}=\rho_{\phi,N_i}a^3,\quad\tilde{N}_1=n_1 a^3=\frac{\rho_{N_1}}{M_1}a^3,\quad\tilde{N}_\text{B-L}=n_\text{B-L}a^3,\quad R=\rho_Ra^4,\quad S=s\,a^3
\end{equation}
where $a$ is the scale factor of the Universe. Note that $E_{N_1}, \tilde{N}_1$ become comoving quantities only after RHNs become non-relativistic, i.e. at temperatures below $T_{\rm NR}$ defined in Eq.~\eqref{eq:TNR}. Instead of time $t$, we write the Boltzmann equations in terms of the ratio of scale factor $a$ to its initial value $a_I$, given by,
\begin{equation}
    \label{eq:y}
    y=\frac{a}{a_I}
\end{equation}
We take $a_I=1$ without loss of generality. The expansion rate can be written as,
\begin{equation}
    H=\sqrt{\frac{8\pi(a_I E_\phi\, y+a_I E_{N_1}\, y+R)}{3M_\text{pl}^2\,a_I^4\,y^4}}
\end{equation}
The temperature can be written as,
\begin{equation}
    \label{eq:T}
    T=\left[\frac{30\,R}{\pi^2g_\ast a_I^4y^4}\right]^\frac{1}{4}
\end{equation}
The redshift parameter $z$ is defined in terms of $M_1$,
\begin{equation}
    \label{eq:z}
    z=\frac{M_1}{T}=M_1a_I\left[\frac{\pi^2g_\ast}{30\,R}\right]^\frac{1}{4}y
\end{equation}
The equilibrium energy densities are given by  \cite{Hahn-Woernle:2008tsk,Hahn-Woernle:2009jyb},
\begin{align}
    \nonumber E_{N_1}^{eq} &= \rho_{N_1}^{eq}a^3=\frac{a_I^3M_1^4y^3}{\pi^2}\left[\frac{3}{z^2}K_2(z)+\frac{1}{z}K_1(z)\right],\quad\tilde{N}_1^{eq}=\frac{E_{N_1}^{eq}}{M_1},\\
    n_{N_1}^{eq}&=\frac{E_{N_1}^{eq}}{M_1 y^3},\quad n_R^{eq}=\frac{3\zeta(3)}{4\pi^2}\times2\left(\frac{M_1}{z(y)}\right)^3
\end{align}
The inverse decay rate can be written as ~\cite{Buchmuller:2004nz},
\begin{equation}
    \Gamma_{ID}=\frac{1}{2}\Gamma_{N_1}\frac{n_{N_1}^{eq}}{n_R^{eq}}
\end{equation}
As described earlier, we ignore the inverse decay term of $\phi$ as well as the RHN scattering terms involving $\Gamma_{\rm scatt}$ and $\Gamma_{ID}^{\rm scatt}$ in the Boltzmann equations given in Eq. \eqref{eq:boltzrho}, and rewrite them as
\begin{align}
\label{eq:boltzE}
\nonumber\frac{d E_{\phi}}{d y} & =-\frac{ \Gamma_{\phi}}{\mathcal{H}}\frac{E_\phi}{y}, \\
\nonumber\frac{d E_{N_1}}{d y} & =\frac{\Gamma_{\phi\rightarrow N_1N_1}}{\mathcal{H}} \frac{E_{\phi}}{y}-\frac{\Gamma_{N_1}}{\mathcal{H} y}\left(E_{N_1}-E_{N_1}^{e q}\right) - \frac{E_{N_1}}{y}\Theta(T-T_{\rm NR}), \\
\nonumber\frac{d \tilde{N}_{B-L}}{d y} & =-\frac{\Gamma_{N_1}}{\mathcal{H} y}\left[\epsilon\left(\tilde{N}_1-\tilde{N}^{eq}_1\right)+\frac{1}{2}\frac{n_{N_1}^{eq}}{n_R^{eq}} \tilde{N}_{B-L}\right],\\
\nonumber\frac{d R}{d y} & = \frac{ \Gamma_{\phi\rightarrow R}}{\mathcal{H}}E_\phi+\frac{\Gamma_{N_1}}{\mathcal{H}}\left(E_{N_1}-E_{N_1}^{eq}\right)
\end{align}
where $\Theta(x)$ is the Heaviside step function separating the relativistic stage of RHNs from the non-relativistic stage,
\begin{equation}
    \Theta(x)=\begin{cases}
        1 & \text{for } x>1\\
        0 & \text{otherwise}
    \end{cases}
\end{equation}
By solving the above equations, one can calculate the efficiency parameter to be,
\begin{equation}
    \kappa_f=-\frac{4}{3}\epsilon^{-1}R^{-3/4}\tilde{N}_\text{B-L}\left[\frac{\pi^4g_\ast^{3/4}}{30^3\zeta(3)^4}\right]
\end{equation}


\vspace{10em}
\newpage


\bibliographystyle{JHEP}
\bibliography{ref}
\end{document}